\documentclass[preprint,12pt,authoryear]{elsarticle}

\usepackage{amssymb}
\usepackage{placeins}
\usepackage{float}
\usepackage{tabularx}

\journal{Progress in Oceanography}

\begin{document}

\begin{frontmatter}



\title{Internal tides in the Mediterranean Sea}


\author[CMCC,UniBo]{Bethany McDonagh}
\author[MPIM,UHH]{Jin-Song von Storch}
\author[CMCC]{Emanuela Clementi}
\author[UniBo]{Nadia Pinardi}

\affiliation[CMCC]{organization={CMCC Foundation – Euro-Mediterranean Center on Climate Change},
            country={Italy}}

\affiliation[UniBo]{organization={Department of Physics and Astronomy, University of Bologna},
            city={Bologna},
            country={Italy}}

\affiliation[MPIM]{organization={Max Planck Institute for Meteorology},
            city={Hamburg},
            country={Germany}}

\affiliation[UHH]{organization={Center for Earth System Research and Sustainability (CEN), University of Hamburg},
            city={Hamburg},
            country={Germany}}

\begin{abstract}
The generation and propagation sites of internal tides in the Mediterranean Sea are mapped through a comprehensive high-resolution numerical study. Two ocean general circulation models were used for this: NEMO v3.6, and ICON-O, both hydrostatic ocean models based on primitive equations with Boussinesq approximation, where NEMO is a regional Mediterranean Sea model with an Atlantic box, and ICON a global model. Internal tides are widespread in the Mediterranean Sea. The primary generation sites: the Gibraltar Strait, Sicily Strait/Malta Bank, and Hellenic Arc, are mapped through analysis of the tidal barotropic to baroclinic energy conversion. Semidiurnal internal tides can propagate for hundreds of kilometres from these generation sites into the Algerian Sea, Tyrrhenian Sea, and Ionian Sea respectively. Diurnal internal tides remain trapped along the bathymetry, and are generated in the central Mediterranean Sea and southeastern coasts of the basin. The total energy used for internal tide generation in the Mediterranean Sea is 2.89 GW in NEMO and 1.36 GW in ICON. Wavelengths of the first baroclinic modes of the M2 tide are calculated in various regions of the Mediterranean Sea where internal tides are propagating, comparing model outputs to a theory-based calculation. The models are also intercompared to investigate the differences between them in their representation of internal tides. 

\end{abstract}


\begin{highlights}
\item Internal tides are widespread in the Mediterranean Sea
\item They are generated in regions including the Gibraltar Strait, Sicily Strait, and Hellenic Arc
\item Semidiurnal internal tides are able to propagate for hundreds of kilometres
\end{highlights}


\end{frontmatter}


\section{Introduction}
\label{introduction}

Internal, or baroclinic, tides are internal waves at tidal frequencies, which are generated when barotropic tides interact with topography in the ocean, often at shelf breaks \cite[]{KL16}, ocean ridges \cite[]{MHJ01, NH01}, and in narrow straits \cite[]{MTV02, BKL14}. These waves can dissipate close to their generation sites in the case of high mode internal tides, or propagate away for up to thousands of kilometres, in the case of low-mode, superinertial waves \cite[]{A22}. Internal tides are responsible for around 1.0 TW of energy dissipation in the global ocean \cite[]{ER03}, meaning that understanding their sites of generation, propagation, and energy dissipation is of great importance when detailing the energy budget of the global ocean. Internal tides also contribute to deep ocean mixing \cite[]{MW98}, so they need to be correctly represented in numerical ocean general circulation models.

In many modern ocean general circulation models, the dissipation of energy from internal tides is represented with a parameterization of tidal mixing, commonly that of \cite{SSJ02}. However, these parameterizations only represent a fraction of the total internal tidal energy dissipation, in particular that of high-mode internal tides which dissipate at their generation sites. The remaining energy, which relates to the low modes of superinertial internal tides, is propagated away. In the Mediterranean Sea, due to its latitude range of 30$^\circ$N-46$^\circ$N, diurnal internal tides have frequencies lower than the inertial frequency (in the Mediterranean Sea, this varies according to latitude between 24 hours in the south and 16.7 hours in the north) and remain trapped along the bathymetry, while semidiurnal internal tides are superinertial and are able to freely propagate away from their generation sites, so both the dissipation of internal tidal kinetic energy close to the generation site and wave propagation over long distances need to be considered in the model’s mixing parameterizations. 

In most of the Mediterranean Sea, the amplitude of barotropic tides is lower than in many other regions of the global ocean, except at the Gibraltar Strait, Gulf of Gabes, and the North Adriatic Sea \cite[]{TPF95, MCG23}. However, regional studies within the Mediterranean Sea such as \cite{MTV02}, \cite{OPF23}, and \cite{AGZ12} have indicated that internal tides are generated in several regions of the Mediterranean Sea, notwithstanding the low amplitude barotropic tidal energy. Additionally, These works suggest that a basin-wide study is needed. 

The Gibraltar Strait is a region in which internal tides have been studied in more detail, through both observational and modelling methods. \cite{MTV02} found that semidiurnal vertical oscillations due to internal tide generation at the Camarinal Sill have amplitudes exceeding 200m, and that this energy causes turbulent mixing in the Gibraltar Strait. Internal tides in the Gibraltar Strait contribute to the bottom pressure \cite[]{CWR90}, and propagate in both directions from the Camarinal Sill \cite[]{LVP00}.

\begin{figure}[]
\centering
\includegraphics[width=\linewidth, trim=0cm 0cm 0cm 0cm, clip]{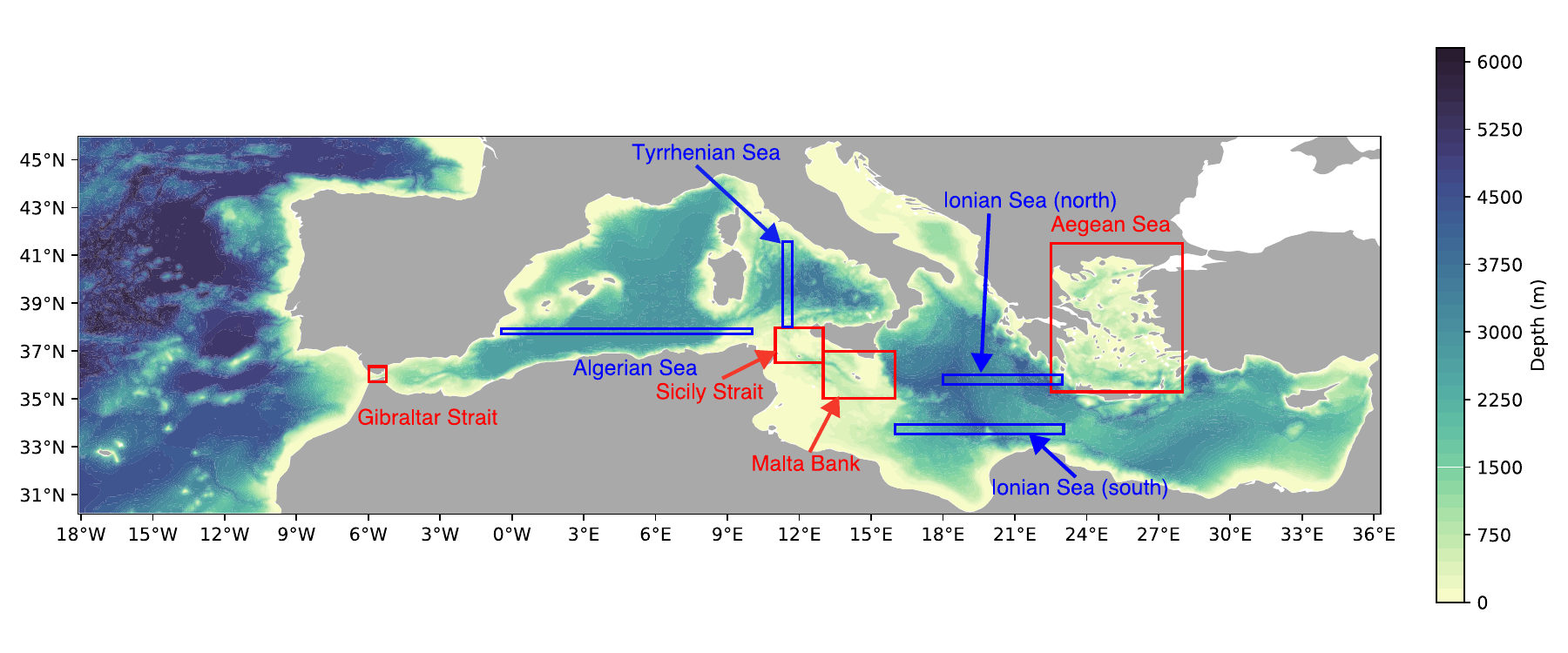}
\caption{Map of NEMO model domain with model bathymetry in coloured contours. Key regions from previous literature for internal tides studies are highlighted in red boxes, and further regions of study in this work are in blue boxes (see \ref{extraregions}).}
\label{fig:domain_map}
\end{figure}

Studies such as \cite{GSA04} and \cite{OPF23} investigated internal tides in the central Mediterranean Sea, finding that internal tides are generated over the complex topography in both the Sicily Strait \cite[]{GSA04}, and Malta Bank \cite[]{OPF23}, affecting the dynamics in both regions. Other observational studies have been carried out in the Aegean Sea (Alford et al., 2012), where Doppler current profiles were used to identify propagating semidiurnal internal tides, and in the Mid-Adriatic \cite[]{MOP09}, where potential trapped diurnal internal waves around islands were found. These studies have shown that internal tides are generated at several sites in the Mediterranean Sea, but research has often been limited by spatial and temporal availability of observational data, which comes mostly from cruises, and limited-area models. The propagation of internal tides in the whole Mediterranean Sea has not previously been investigated. A summary of the previously studied regions of internal tides within the Mediterranean is shown in Figure \ref{fig:domain_map}. 

In other regions of the ocean, and for the global ocean in its entirety, numerical studies have proven to be valuable in understanding internal tides. Studies including \cite{AGZ12}, \cite{MCF12}, \cite{SAR12}, and \cite{LvSM15, LvSM17} demonstrate the value of using a high-resolution general circulation ocean model to investigate internal tides in the global ocean. This is done through identifying the characteristics of the internal tides and diagnosing the generation and propagation of internal tides. Additionally, analysis of internal tide wavelengths and vertical structure modes are used to understand the limitations of the numerical models. Basin- wide studies of oceans and seas such as \cite{NH01} in the Pacific Ocean, and \cite{KL16} in the Atlantic Ocean show the crucial role of topography in the generation of internal tides. These investigations also demonstrate how this phenomenon integrates into the larger framework of internal tidal energy dynamics, including its interactions with major ocean currents like the Gulf Stream. Internal tides also interact with mesoscale eddies, as discussed in \cite{DL14} and \cite{GWC23}, where eddies change the phase speeds and cause refraction of internal tides. 

In this study, we carried out an analysis of numerical simulations to understand the dynamics of internal tides in the Mediterranean Sea, with particular emphasis on finding generation sites, regions of semidiurnal internal tidal propagation, and the vertical modes of internal tides. Two ocean general circulation models are used, each with hourly outputs over a one-month period. Section \ref{models} details the modelling framework and the spectral analysis used throughout the work, Section \ref{results} contains results from the model outputs, Section \ref{differences} discusses the key differences between the two models, and Section \ref{conclusion} provides a conclusion. An intercomparison of tides in the models is available in \ref{intercomparison}, and results for additional regions are shown in \ref{extraregions}. 

\section{Model configurations}
\label{models}

\subsection{NEMO}

The implementation of version 3.6 of the general circulation ocean model NEMO \cite[]{MDI98} in the Mediterranean Sea is used, based on the operational models described in \cite{CMEMS_EAS6} and \cite{CCC23}, and the NEMO model domain is shown in Figure \ref{fig:domain_map}. This NEMO configuration has a horizontal resolution of $\frac{1}{24}^\circ$, and 141 uneven z* vertical layers with partial steps at the bottom. Lateral open boundary conditions are used in the Atlantic Ocean from the Copernicus Marine global forecast and analysis \cite[]{CMEMS_global}, and in the Dardanelles Strait from a Marmara Sea box model \cite[]{MIL15}. 

The Atlantic Ocean boundary includes tidal components and equilibrium tidal forcing is imposed as a surface forcing with eight tidal equilibrium components (M2, S2, K1, O1, N2, Q1, K2, P1) from TPXO9 \cite[]{EE02}, which are described in Table \ref{tab:tidal_components}. 

\begin{table}[]
    \centering
    \begin{tabular}{|c|c|c|}
        \hline
        \textbf{Tidal component} & \textbf{Period (hours)} & \textbf{Description} \\
        \hline
        M2 & 12.421 & Principal lunar semidiurnal \\ & & tidal constituent \\
        S2 & 12.000 & Principal solar semidiurnal \\ & & tidal constituent \\
        K1 & 23.934 & Lunisolar diurnal tidal \\ & & constituent \\
        N2 & 12.658 & Larger lunar elliptic semidiurnal \\ & & tidal constituent \\
        P1 & 24.066 & Solar diurnal tidal \\ & & constituent \\
        Q1 & 26.868 & Larger lunar elliptic diurnal \\ & & tidal constituent \\
        K2 & 11.967 & Lunisolar semidiurnal tidal \\ & & constituent \\
        \hline
    \end{tabular}
    \caption{Tidal constituent components used in the model for this work, with their respective periods and astronomical descriptions.}
    \label{tab:tidal_components}
\end{table}

The atmospheric forcing for the one-month period in 2022 analysed in this work starts from the six-hourly temporal resolution European Centre for Medium-Range Weather Forecasts (ECMWF) analyses for surface atmospheric variables, which are used by bulk formulae from \cite{PLP10} to calculate water, momentum, and heat fluxes. Furthermore, monthly mean climatologies for the 39 rivers with the largest water flux are included in the surface layer. These are from \cite{FVG99} for the Ebro, Nile, Po, and Rhône rivers, \cite{R96} for the Vjosë and Seman rivers, \cite{D96} for the Buna and Bojana rivers, and \cite{P12} for the remaining 32 rivers. Initial conditions of temperature and salinity are from the 2005-2012 winter climatology of \cite{WORLD_OCEAN_ATLAS}, and the model was integrated from January 2015, so has an extensive spin-up period. 

The model bathymetry is provided by \cite{GEBCO}, bilinearly interpolated onto the model grid, with some additional changes in several locations. Firstly, two points in the Bay of Biscay are closed and several points along the Croatian coastline are modified for stability purposes, to avoid spurious tidal generation around complex bathymetry. These modifications follow \cite{CMEMS_EAS6}. Furthermore, several points in the Gibraltar Strait are modified to improve the mass transport values and avoid spurious vertical mixing in the Gibraltar Strait region. These changes are made according to \cite{M24}. Vertical mixing in the model configuration uses a turbulent kinetic energy (TKE) closure scheme, which is again tuned following \cite{M24}. 

\subsection{ICON-O}

To provide a robust numerical analysis of internal tides in the Mediterranean Sea, a second dataset, derived from a global scale implementation of the ICON-O \cite[]{KBJ22} primitive equations general circulation model, is used. 

ICON-O uses an icosahedral grid, which splits the Earth’s surface initially into twenty equilateral triangles and then bisects these triangles further for higher resolution simulations, resulting in so- called R2B\textit{n} resolutions, where \textit{n} is the number of times that the original icosahedron is bisected \cite[]{GBC18}. The simulation of ICON-O used in this work has a resolution of R2B9, meaning that the initial grid is divided nine further times, which is approximately equivalent to a horizontal resolution of 5 km when interpolated onto a regular latitude-longitude grid. This resolution is slightly lower than, but comparable to, NEMO’s $\frac{1}{24}^\circ$ horizontal resolution in the Mediterranean Sea. ICON-O has 128 z* vertical levels without partial cells at the bottom, and uses the SRTM30 bathymetry \cite[]{SRTM30_bathy} interpolated onto the model grid. Narrow straits (Gibraltar, Dardanelles, and Bosphorus in the Mediterranean/Black Sea region) are manually checked to ensure that they are open, but no further modifications are made. The Black and Marmara seas are included in the ICON-O model, so no boundary condition is required at the Dardanelles Strait.

The world’s fifty largest rivers are included in the configuration \cite[]{GHG93}, including four that flow into the Mediterranean Sea: the Ebro, Nile, Po, and Rhône. These use flux values from the daily climatology by \cite{R06}. Atmospheric forcing is from ERA5 \cite[]{ERA5} with a temporal frequency of one hour, at a resolution of 30 km interpolated onto the model grid. 

Tides are included in ICON-O via a lunisolar gravitational term in the momentum equation, which is the horizontal component of the difference between the gravitational forces of the Sun and Moon and the centrifugal force due to the rotation of the Earth \cite[]{LLK21}. Therefore ICON-O includes all tidal components. The accuracy of tides in ICON at lower resolutions (R2B6 to R2B8) was assessed in \cite{vSHL23}.

Vertical mixing in ICON uses a TKE scheme also based on \cite{GGL90} and has a similar configuration to NEMO for this parameterization. The model uses further parameterizations of secondary tidal processes: topographic wave drag, and self-attraction and loading, which are not included in the NEMO configuration. Some of the key differences between the two model configurations are summarised in Table \ref{tab:NEMO_ICON}.

\begin{table}[]
    \centering
    \begin{tabularx}{\linewidth}{|X|X|X|}
    \hline
    \textbf{Model component} & \textbf{NEMO} & \textbf{ICON} \\
    \hline
    Region & Mediterranean Sea with Atlantic Box & Global \\
    Grid & Rectangular latitude-longitude & Icosahedral \\
    Horizontal resolution & $\frac{1}{24}^\circ$ (3.8km) & R2B9 (around 5km) \\
    Vertical resolution & 141 z* levels with partial cells & 128 z* levels without partial cells \\
    Tides & 8 tidal equilibrium tidal forcing components (M2, S2, K1, O1, N2, Q1, K2, P1) in the momentum and at the Atlantic lateral boundary & Full lunisolar tidal potential \\
    Secondary tidal processes & None & Topographic wave drag and self attraction and loading \\
    Rivers & 39 rivers from climatology & 4 Mediterranean rivers from climatology \\
    Bathymetry & GEBCO (30") & SRTM30 (30") \\
    \hline
    \end{tabularx}
    \caption{Comparison of some key features of the NEMO and ICON implementations used in this work.}
    \label{tab:NEMO_ICON}
\end{table} 

One month (March 2022) of data is analysed for each model, with hourly outputs of three-dimensional currents, temperature, salinity, Brunt-Väisälä frequency, and two-dimensional sea surface height. 

\section{Results}
\label{results}

\subsection{Internal tide generation}

Internal tide generation can be quantified in terms of $C$, the conversion from barotropic tidal energy to internal tide energy \cite[]{KF12, M13, NH14, LvS20}, defined as: 

\begin{equation}
    C = \int_{-D}^\eta g \rho' W dz 
\end{equation}

where $\rho'$ is the density perturbation associated with an internal tide, $W$ is the tidal vertical velocity, $\eta$ is the sea surface height, and $D$ is the depth of the ocean. $W$ is obtained from:

\begin{equation}
    W(z) = -\nabla (D + z) U = U \cdot \nabla D + (D + z)\cdot \nabla U
\end{equation}

where $U$ is the barotropic horizontal tidal velocity, derived from a harmonic analysis of the horizontal currents, based on the method of \cite{FCB09}.

To derive $\rho'$ associated with an internal tide, the full density is decomposed into:

\begin{equation}
\label{rho_decomposition}
    \rho(x, y, z, t) = \rho_0 + \rho_b(x, y, z) + \rho^{*}(x, y, z, t)
\end{equation}

where $\rho_0$ is the constant reference density, and $\rho_b$ is the background time-mean density at each grid point, so that $\rho_0 + \rho_b$ is the total background mean density. A harmonic analysis on the time varying $\rho^*$ in Eq. \ref{rho_decomposition} is carried out to obtain the density perturbation $\rho'$ associated with internal tides at different tidal frequencies.  

\begin{figure}[]
\centering
\includegraphics[width=\linewidth, trim=0cm 0cm 0cm 0cm, clip]{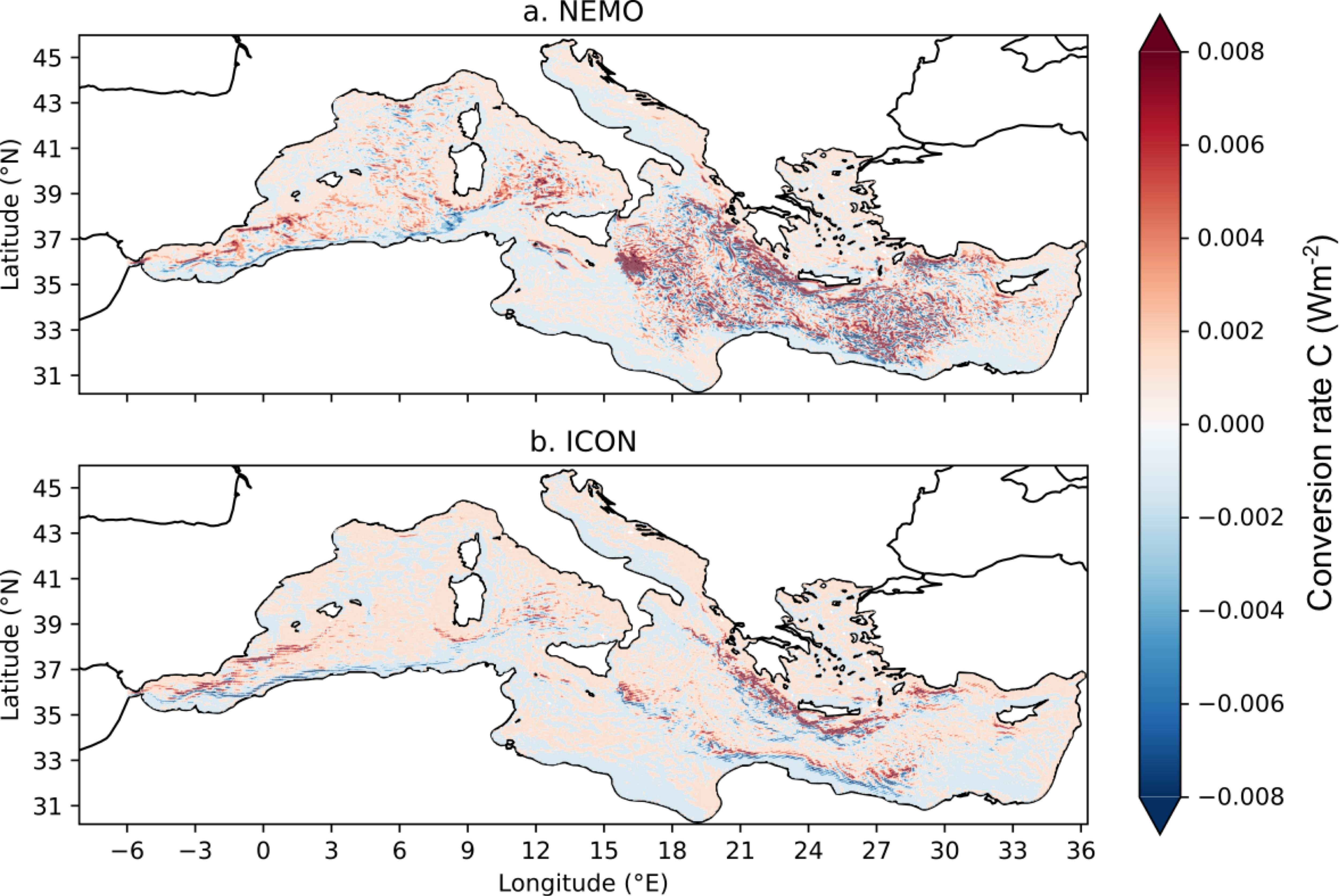}
\caption{Maps of the barotropic to baroclinic energy conversion term ($C$) in the Mediterranean Sea, for a. NEMO, and b. ICON.}
\label{fig:conversion_all}
\end{figure}

Once $C$ is calculated using the values of $u$, $v$, and $\rho'$ for each of the eight tidal components included in the NEMO experiment, it is summed to find a total value for $C$. Positive values of $C$ indicate the generation of internal tides by barotropic tides. Figure \ref{fig:conversion_all} shows a map of $C$ for the Mediterranean Sea in the two experiments. These maps confirm and quantify the generation sites of internal tides. 

\begin{figure}[]
\centering
\includegraphics[width=\linewidth, trim=0cm 0cm 0cm 0cm, clip]{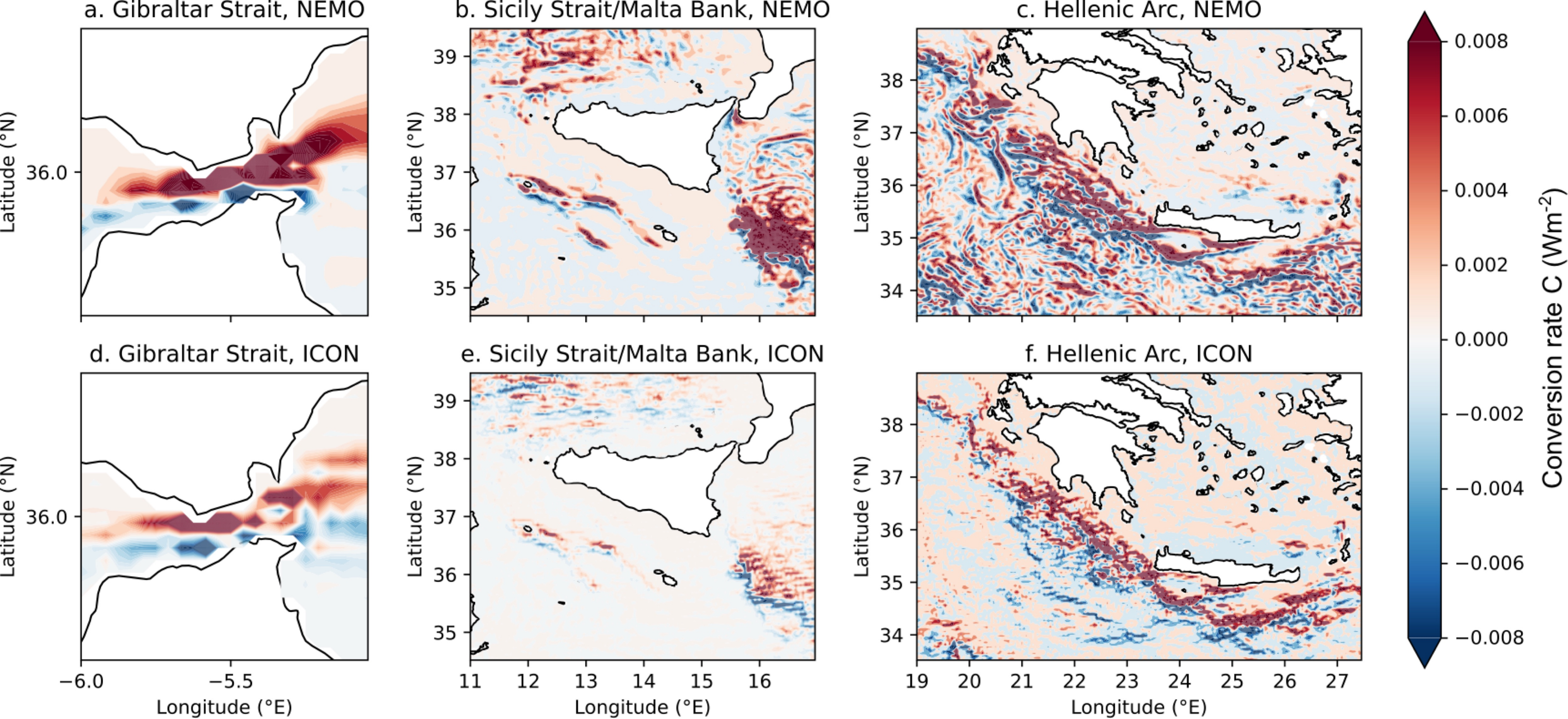}
\caption{Maps of the barotropic to baroclinic energy conversion term ($C$) in key regions of the Mediterranean Sea: the Gibraltar Strait/Alboran Sea (a, d), the Sicily Strait/Malta Bank (b, e), and the Hellenic Arc region (c, f).}
\label{fig:conversion_regions}
\end{figure}

Barotropic to baroclinic energy conversion for internal tide generation takes place in a wide variety of regions, but most notably in the Gibraltar Strait/Alboran Sea, the Algerian Sea, the Tyrrhenian Sea, the Sicily Strait/Malta Bank, the Ionian Sea, and in NEMO, the eastern and north-western Mediterranean Sea  in general. Looking at these regions more closely in Figure \ref{fig:conversion_regions}, we can see that the narrow and steep Gibraltar Strait (Figure \ref{fig:conversion_regions} a, d) is a major generator of baroclinic energy, where tidal velocity is high. The Hellenic Arc (Figure \ref{fig:conversion_regions} c, f) is the next most significant region for generation of internal wave kinetic energy, with large values of the conversion term $C$ particularly widespread in the NEMO experiment. The narrow passage in the Sicily Strait and the steep slope in the Malta Bank (Figure \ref{fig:conversion_regions} b, e) also have large values of $C$. The higher values of $C$ in NEMO in the Malta Bank likely come from the stronger baroclinic K1, as K1 is the predominant internal tide component in this region \cite[]{OPF23}. The region-averaged value of $C$ in Fig. \ref{fig:domain_map} regions is positive for both models with larger values for NEMO. 

\begin{figure}[]
\centering
\includegraphics[width=\linewidth, trim=0cm 0cm 0cm 0cm, clip]{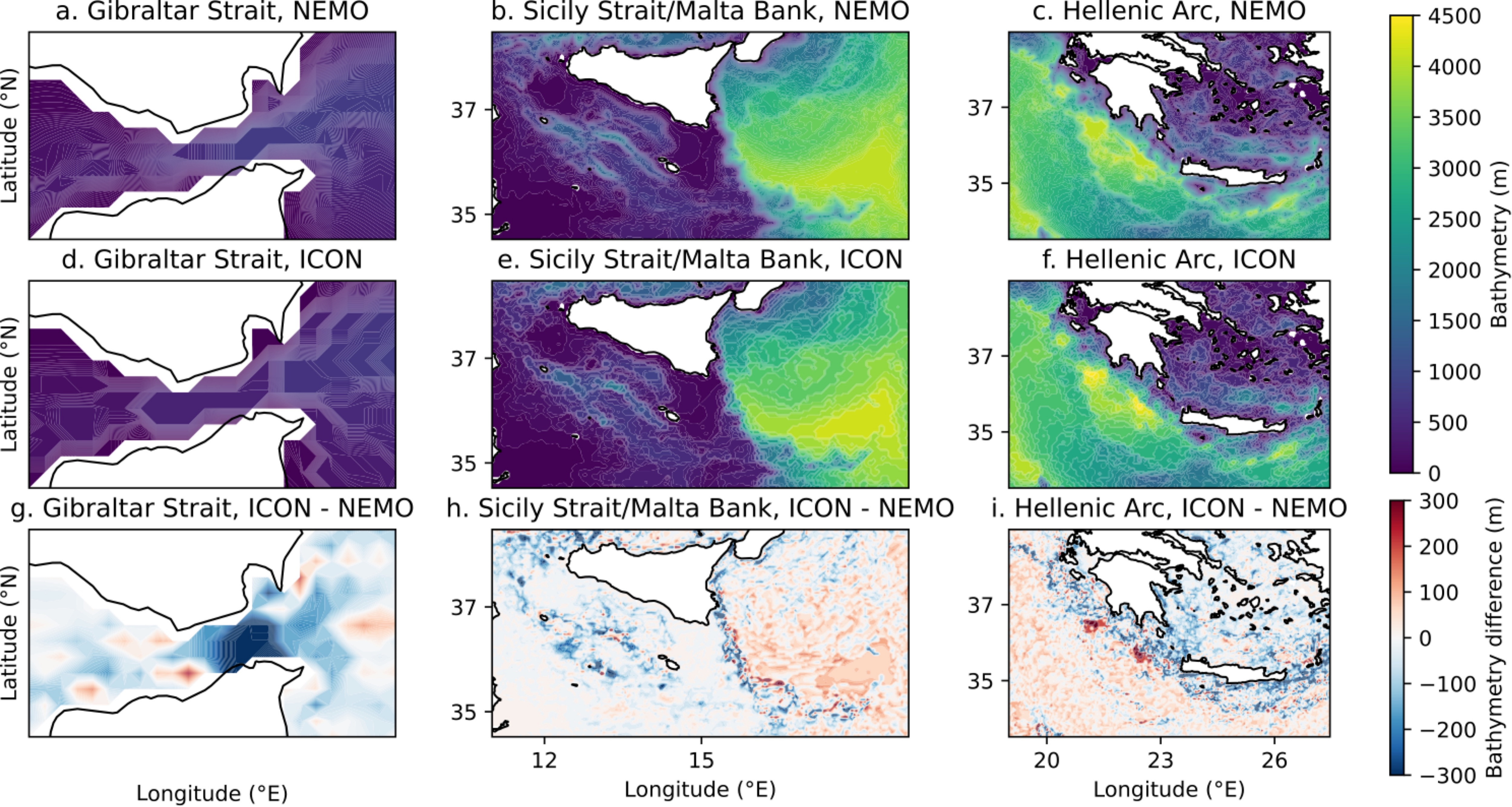}
\caption{Model bathymetry in key regions for internal tide generation: Gibraltar Strait/Alboran Sea (a, d, g), Sicily Strait/Malta Bank (b, e, h), and Hellenic Arc (c, f, i), for NEMO (a, b, c), ICON (d, e, f), and the difference between models, ICON - NEMO (g, h, i).}
\label{fig:bathy_diff_regions}
\end{figure}

Some of the differences in $C$ between the two models is likely connected to the different topography in some of the key internal tide generation regions, as is shown in Figure \ref{fig:bathy_diff_regions}. In the narrowest part of the Gibraltar Strait, the most important region in the Mediterranean Sea for internal tide generation, the bathymetry is deeper in the NEMO implementation than it is in that of ICON. This is likely to affect the generation of internal tides and their propagation throughout the western Mediterranean basin. Another key bathymetry difference is in the Malta Bank (Figure \ref{fig:bathy_diff_regions}h): in the ICON implementation the bathymetry is steeper and placed differently in comparison to the NEMO experiment, and the Ionian Sea is deeper while the Sicily Strait passage is shallower in the ICON experiment compared to that of NEMO. Finally, the deeper Ionian Sea also affects the Hellenic Arc region (Figure \ref{fig:bathy_diff_regions}i). 

\begin{figure}[]
\centering
\includegraphics[width=\linewidth, trim=0cm 0cm 0cm 0cm, clip]{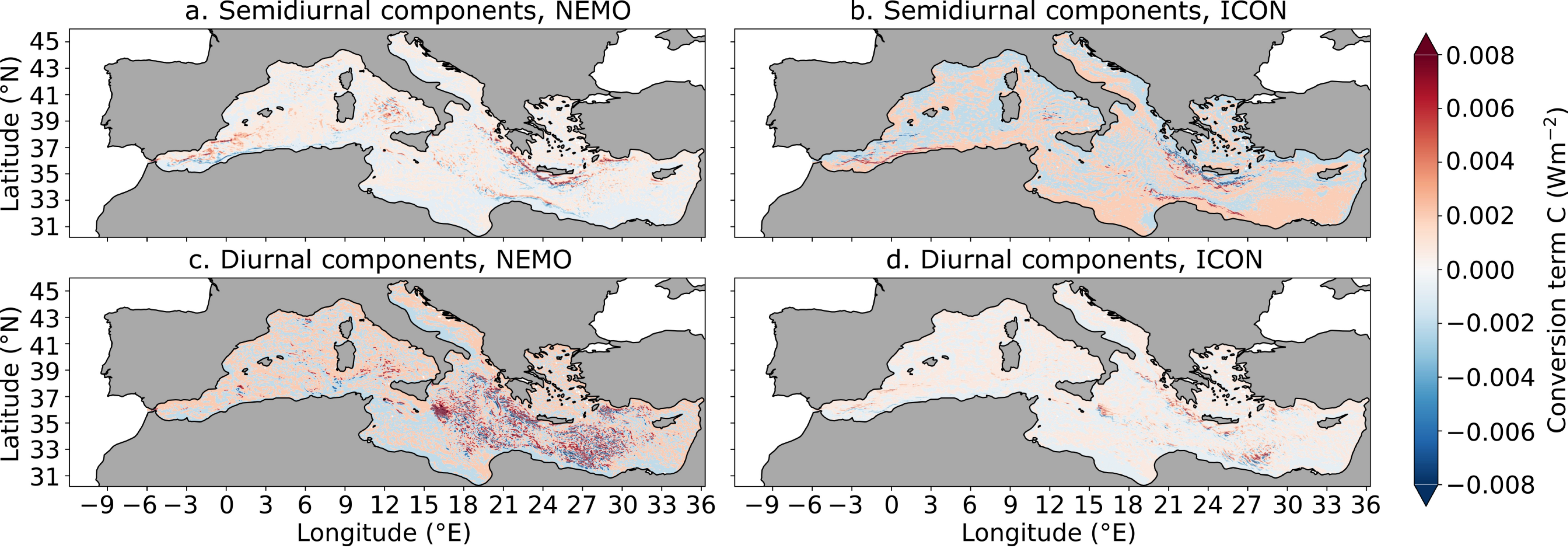}
\caption{Maps of the barotropic to baroclinic energy conversion term ($C$) in the Mediterranean Sea, split into semidiurnal (M2, S2, K2, and N2; a. NEMO and b. ICON) and diurnal (K1, O1, P1, and Q1; c. NEMO and d. ICON) components.}
\label{fig:conversion_components}
\end{figure}

It is also valuable to compare the generation of internal tides between diurnal and semidiurnal components, since we know that semidiurnal internal tides can propagate freely in the Mediterranean Sea while diurnal internal tides remain trapped along the bathymetry. Figure \ref{fig:conversion_components} shows the maps from Fig. \ref{fig:conversion_all} split into diurnal and semidiurnal components. The regions with high values of $C$ are similar for the semidiurnal components in the two experiments, with the key generation sites in the Gibraltar Strait and Hellenic Arc prominent in both models, while the diurnal components are much more widespread in the Eastern Mediterranean in NEMO than they are in ICON: we argue that the diurnal internal tide generation is the primary reason for the model differences. In NEMO, internal tide generation occurs over large swathes of the Eastern Mediterranean, with a strong generation site at the Malta Bank, as expected from \cite{OPF23}.

It should be noted that there are some large-scale negative values for $C$ for semidiurnal components in ICON-0 and for diurnal components in NEMO. Furthermore, $C$ reaches relatively large negative values in the regions of internal tide generation. This happens when internal tides generated at a site interfere with internal tides generated elsewhere. This can cause $\rho'$ and $W$ to become out of phase which leads to negative values of $C$. The interference can be stronger in a closed basin than in open ocean, due to internal tides reflected at the coasts. We can further quantify internal tide generation by taking an area-weighted mean over each of the regions, although the caveats mentioned above should be considered. To account for the problem of negative values due to interference, only points with positive values are included in the calculation. The area-weighted mean and summed values of the regions in Figure \ref{fig:conversion_regions} are in Table \ref{tab:mean_c}, giving an estimate of the total energy for internal tide generation in the Mediterranean Sea: 2.89 GW in NEMO and 1.36 GW in ICON. The value for the total $C$ in the global ocean is 1.7 TW according to \cite{M13}. The lower value in ICON is due to the smaller generation of diurnal internal tides in the model compared to NEMO. This could be related to differences in the barotropic tides in the model (see \ref{intercomparison}). 

\begin{table}[]
    \centering
    \begin{tabularx}{\linewidth}{|X|X|X|X|X|}
    \hline
    & \multicolumn{2}{c|}{\textbf{Mean (Wm$^{-2}$)}} & \multicolumn{2}{c|}{\textbf{Total (W)}} \\
    \hline
    & \textbf{NEMO} & \textbf{ICON} & \textbf{NEMO} & \textbf{ICON} \\
    \hline
    Mediterran\-ean Sea & 2.24 $\times$ 10$^{-3}$ & 1.06$\times$ 10$^{-3}$ & 2.89$\times$ 10$^{9}$ & 1.36$\times$ 10$^{9}$ \\
    Gibraltar Strait & 4.78$\times$ 10$^{-3}$ & 2.47$\times$ 10$^{-3}$ & 1.26$\times$ 10$^{7}$ & 5.75$\times$ 10$^{6}$ \\
    Sicily Strait/Malta Bank & 2.94$\times$ 10$^{-3}$ & 7.11$\times$ 10$^{-4}$ & 3.94$\times$ 10$^{8}$ & 9.28$\times$ 10$^{7}$ \\
    Hellenic Arc & 3.74$\times$ 10$^{-3}$ & 2.33$\times$ 10$^{-3}$ & 8.00$\times$ 10$^{8}$ & 4.92$\times$ 10$^{8}$ \\
    \hline
    \end{tabularx}
    \caption{Area-weighted mean and area-summed values for $C$ within the regions of Figure \ref{fig:conversion_regions}, in Wm$^{-2}$ for mean values and W for total values.}
    \label{tab:mean_c}
\end{table}

\subsection{Internal tide propagation}

\begin{figure}[]
\centering
\includegraphics[width=\linewidth, trim=0cm 0cm 0cm 0cm, clip]{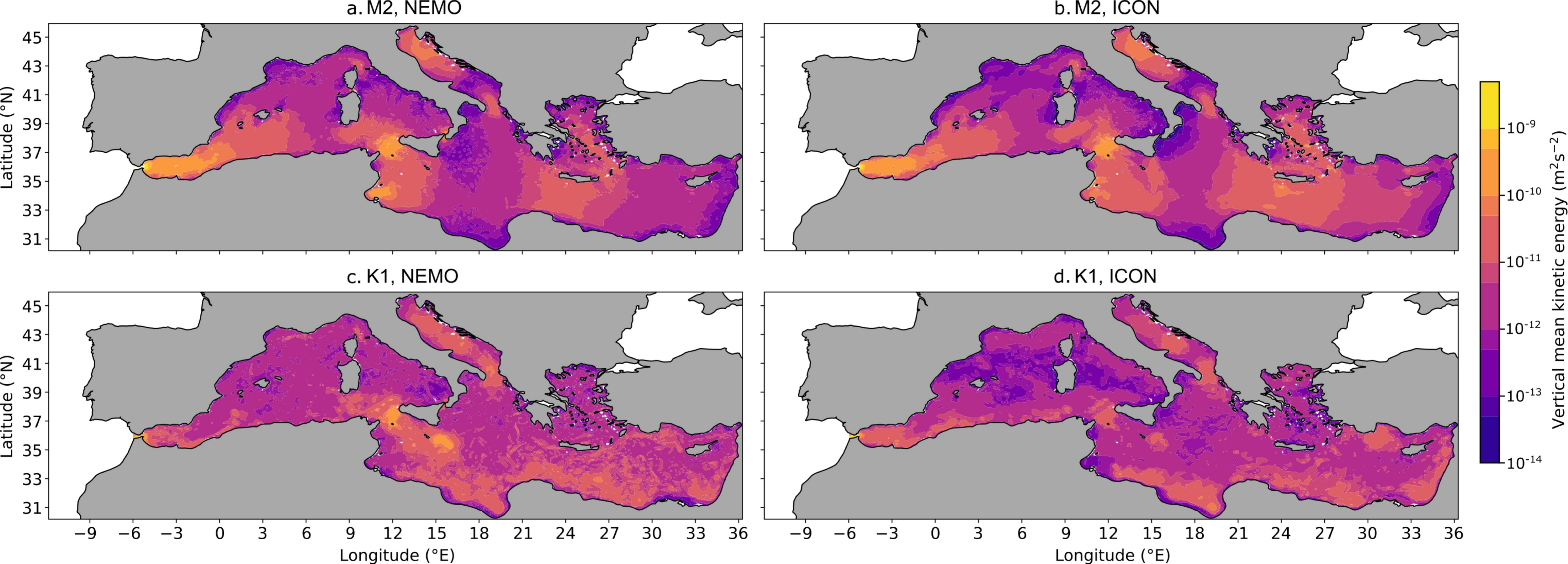}
\caption{Vertical mean baroclinic kinetic energy in the Mediterranean Sea during March 2022, for the M2 component in a. NEMO and b. ICON and for the K1 component in c. NEMO and d. ICON.}
\label{fig:mean_ke}
\end{figure}

A point-by-point harmonic analysis of the three-dimensional horizontal currents over the 744 hours of the model run is carried out in order to isolate tidal currents for each component, using software based on \cite{FCB09}. From this, a vertical mean is removed to calculate the baroclinic component of this as the internal tide kinetic energy for a single layer. Maps of vertical mean baroclinic kinetic energy of the M2 and K1 tidal components in the NEMO and ICON experiments are shown in Figure \ref{fig:mean_ke}. 

The vertical mean baroclinic kinetic energy of the M2 component is similar in the two model implementations. The regions with the highest baroclinic kinetic energy through the water column are similar to those identified in previous work, e.g. \cite{MTV02}; \cite{GSA04}; \cite{AGZ12}: the Gibraltar Strait, Sicily Channel, and Aegean Sea. Moreover, this analysis highlights sites of internal tide generation or propagation that were not mentioned in previous literature: the northern Adriatic Sea, the Corsica Channel and the Cretan Passage. The K1 tidal component has a different distribution of baroclinic kinetic energy in the two models, with particularly high kinetic energy in NEMO in the southern Adriatic Sea and the Malta Bank (as discussed in \cite{OPF23}), as well as along the African coast in the south-east of the basin. The K1 baroclinic kinetic energy is larger in NEMO than in the ICON experiment but it appears in broadly similar regions in the two models.

\begin{figure}[]
\centering
\includegraphics[width=\linewidth, trim=0cm 0cm 0cm 0cm, clip]{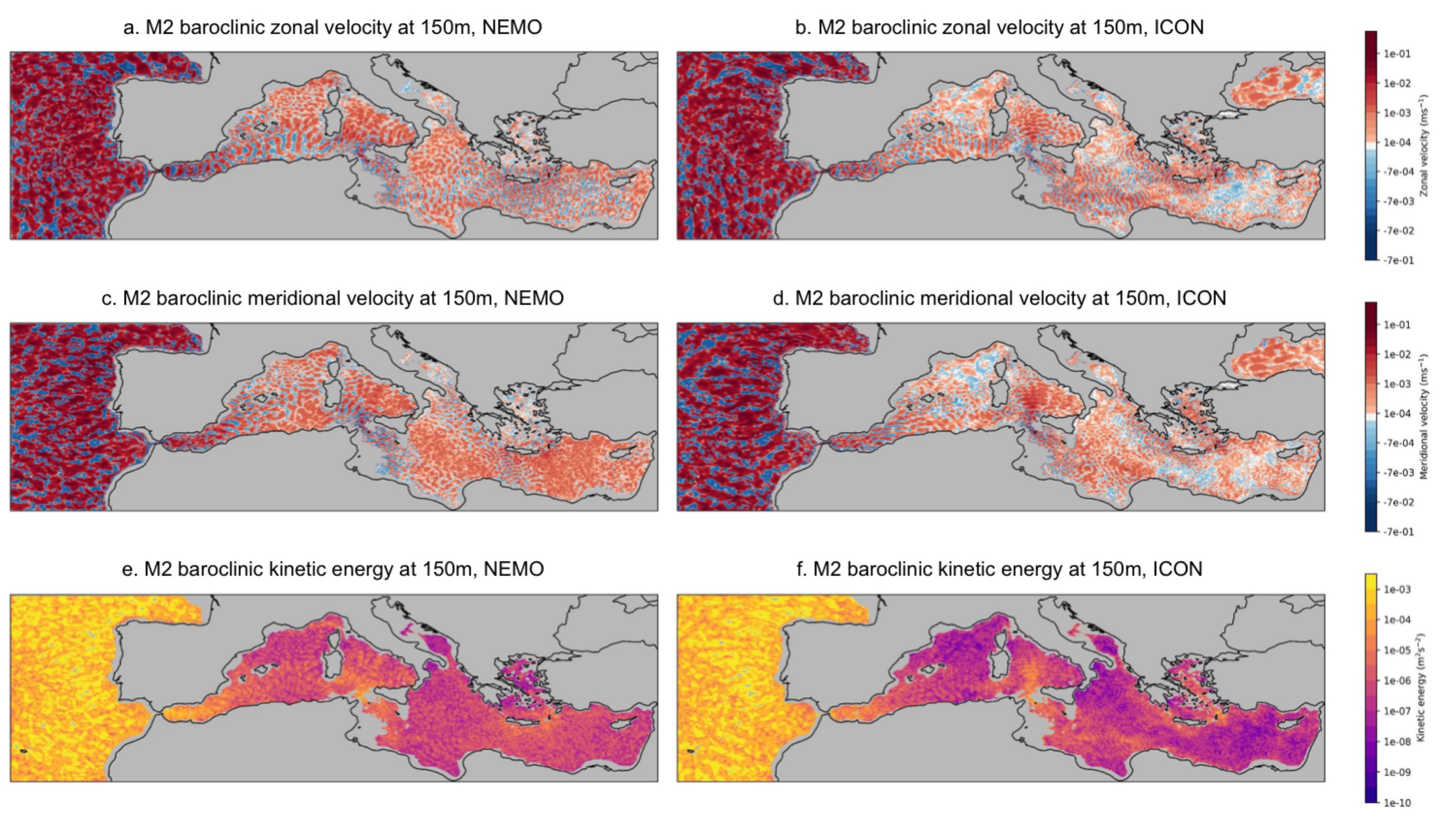}
\caption{Maps of baroclinic M2 current amplitude at the closest vertical level to 150m for a. zonal velocity component in NEMO, b. zonal velocity component in ICON, c. kinetic energy in NEMO, and d. kinetic energy in ICON. }
\label{fig:m2_150}
\end{figure}

Figure \ref{fig:m2_150} shows the baroclinic M2 currents at the vertical level closest to 150m in each model. Only the zonal current is shown, since the meridional currents had similar regions of internal tide propagation. In both models, the propagation of internal tides is visible in the western Mediterranean inflowing current inside the Gibraltar Strait. This current typically flows from the Gibraltar Strait along the north African coast at depths of 0-150m, which has particularly clear tidal beams in the baroclinic zonal velocity (Figures \ref{fig:m2_150}a and b). The waves are also seen in the Tyrrhenian Sea, and in several regions of the eastern Mediterranean, most notably to the west of Crete. There are some key differences between the two experiments: the NEMO experiment shows higher kinetic energy in the western Mediterranean whereas the ICON experiment has more kinetic energy in the Tyrrhenian Sea, but the regions showing propagating waves and tidal beams are broadly similar in the two models. 

\begin{figure}[]
\centering
\includegraphics[width=\linewidth, trim=0cm 0cm 0cm 0cm, clip]{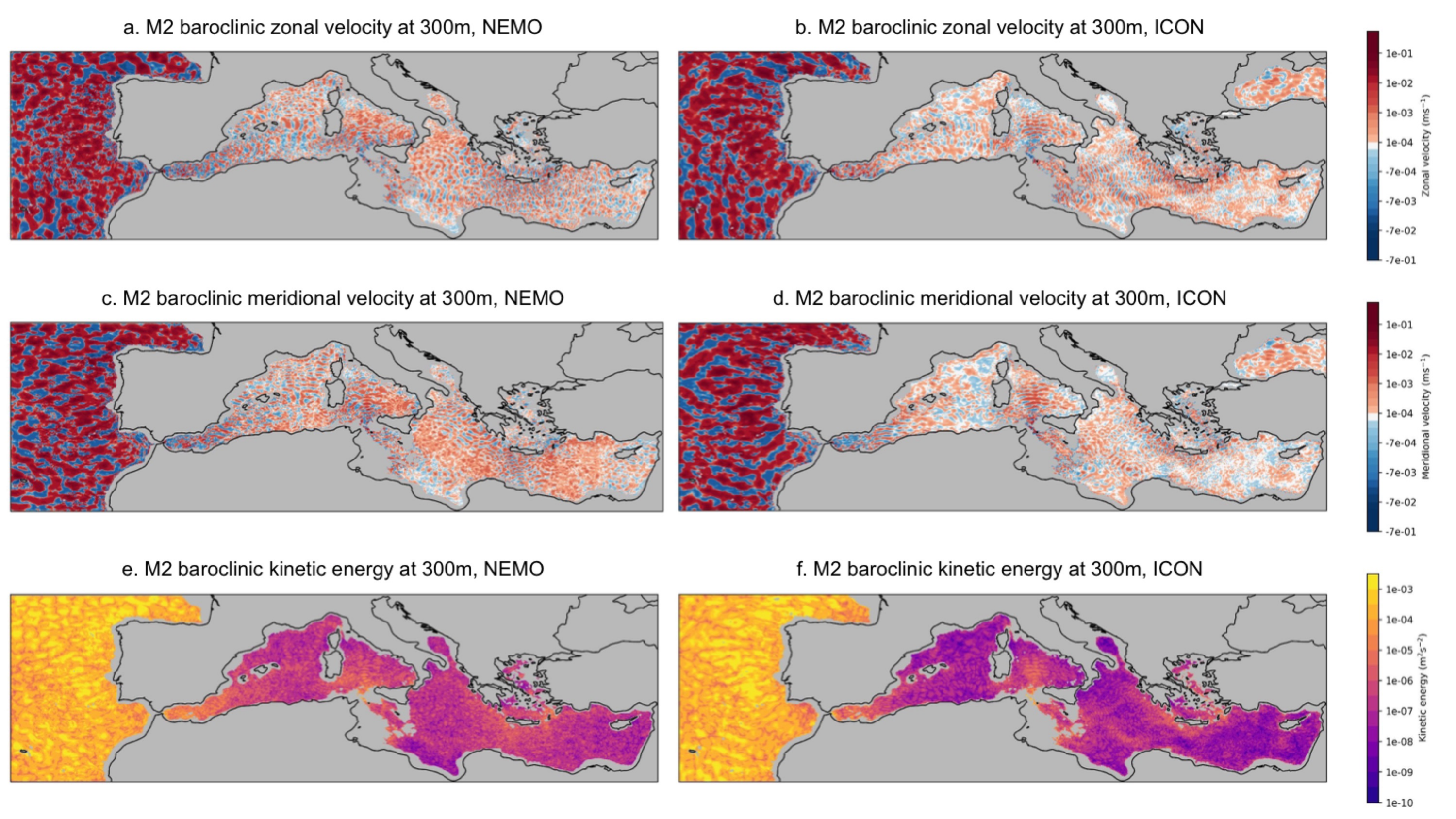}
\caption{Maps of baroclinic M2 current amplitude at the closest vertical level to 300m for a. zonal velocity component in NEMO, b. zonal velocity component in ICON, c. kinetic energy in NEMO, and d. kinetic energy in ICON.}
\label{fig:m2_300}
\end{figure}

\begin{figure}[]
\centering
\includegraphics[width=\linewidth, trim=0cm 0cm 0cm 0cm, clip]{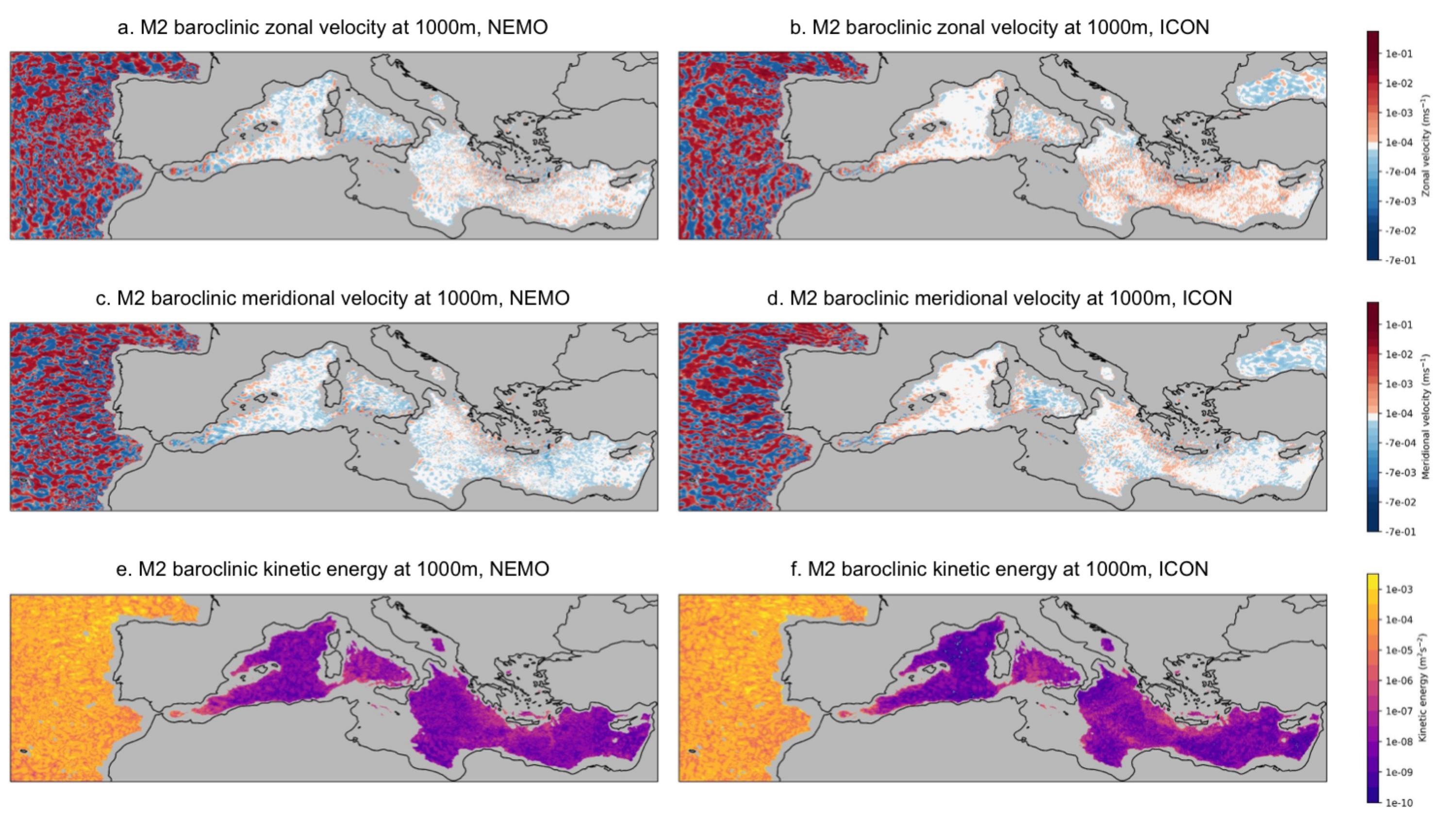}
\caption{Maps of baroclinic M2 current amplitude at the closest vertical level to 1000m for a. zonal velocity component in NEMO, b. zonal velocity component in ICON, c. kinetic energy in NEMO, and d. kinetic energy in ICON.}
\label{fig:m2_1000}
\end{figure}

These waves at the M2 frequency are also visible at deeper layers: Figure \ref{fig:m2_300} shows the same currents at 300m and Figure \ref{fig:m2_1000} at 1000m. At 300m, the kinetic energy is lower than at 150m, but the internal tides are still apparent in the same regions. However, at around 1000m, the structures begin to break down, and tidal beams are no longer present. Since the Sicily Strait is shallower than 1000m at its deepest point, the Mediterranean is split into eastern and western basins, and internal tides are unable to propagate across. In the deeper layers of Figures \ref{fig:m2_300} and \ref{fig:m2_1000}, we see an emergence of baroclinic kinetic energy at the M2 frequency in the Hellenic Arc region, a steep topographic feature between the deep eastern Mediterranean and shallow Aegean Sea. This has not been described before as a site for potential internal tide generation in the eastern Mediterranean Sea, and explains the wave propagation found in the Ionian Sea. 

\begin{figure}[]
\centering
\includegraphics [width=0.8\linewidth, trim=0cm 0cm 0cm 0cm, clip]{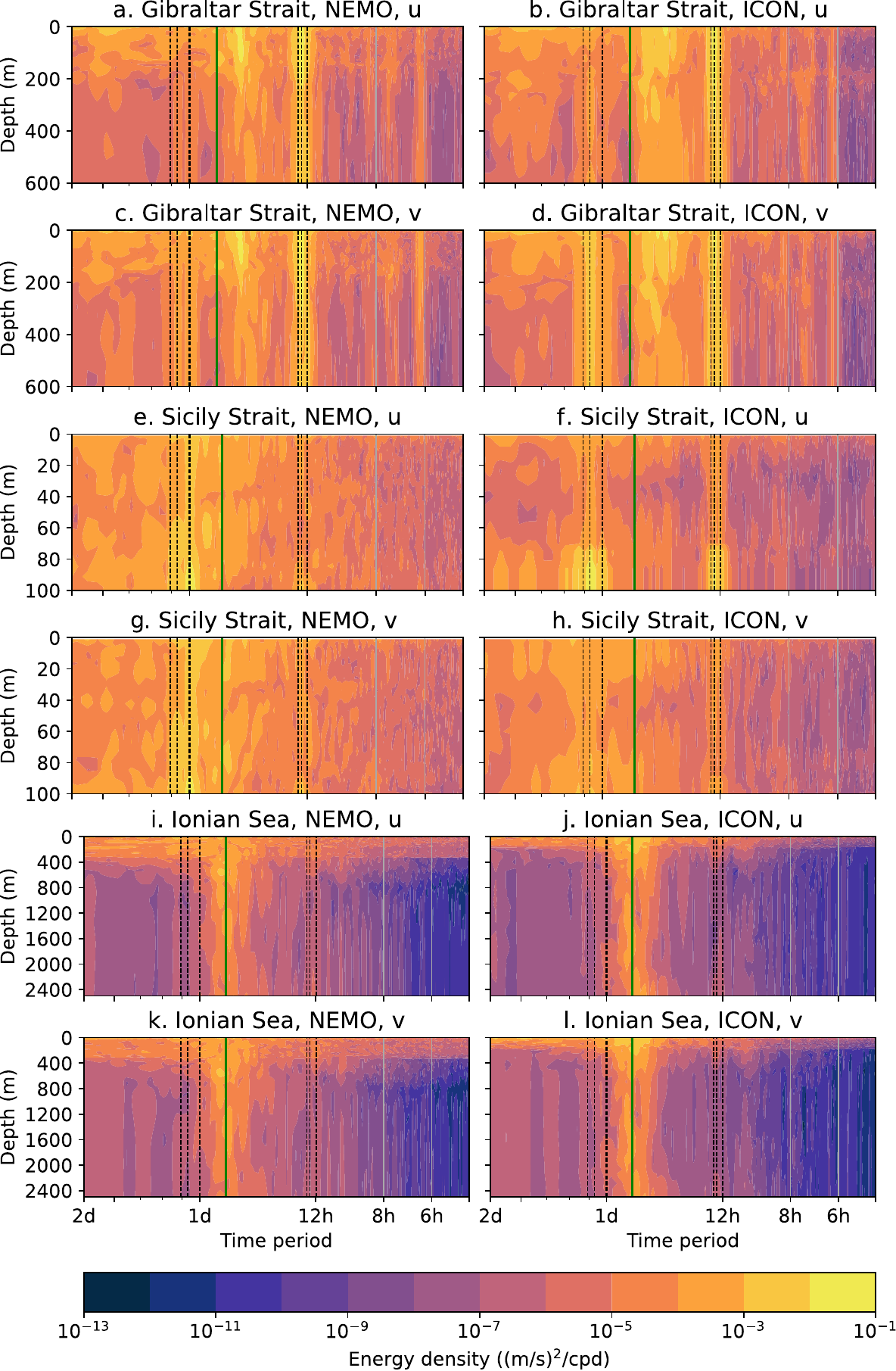}
\caption{Spectra of zonal (u) and meridional (v) currents at three points: the Gibraltar Strait (35.98$^\circ$N, 5.48$^\circ$W) (a-d), Sicily Strait (37.19$^\circ$N, 11.98$^\circ$E) (e-h), and Ionian Sea (35.73$^\circ$N, 20.44$^\circ$E) (i-l), evaluated from NEMO (left) and ICON (right) model simulations through the vertical column. Dashed black lines represent the tidal frequencies included in NEMO and the green line indicates the inertial frequency at the latitude of the point.}
\label{fig:regions_spectra}
\end{figure}

Spectra of the baroclinic horizontal currents throughout the water column are plotted in several key points. These are the centres of the points in Figure \ref{fig:domain_map}, as well as in four further regions, which appear to have internal tide propagation in the maps of Figures \ref{fig:m2_150}-\ref{fig:m2_1000}. These are the Algerian Sea (37.70$^\circ$N, 5.23$^\circ$E), the Tyrrhenian Sea (39.73$^\circ$N, 11.50$^\circ$E), and two points in the Ionian Sea (35.90$^\circ$N, 20.46$^\circ$E and 33.50$^\circ$N, 19.48$^\circ$E), which are the centre points of regions shown in Figure \ref{fig:domain_map}. Three of these eight spectra are shown in Figure \ref{fig:regions_spectra}: the Gibraltar Strait, Sicily Strait, and northern Ionian Sea, while the others can be found in \ref{extraregions}. 

In the Gibraltar Strait (Figure \ref{fig:regions_spectra} a-d), we see high values of energy density at both diurnal and semidiurnal frequencies throughout the water column in both models and for both zonal and meridional currents. In the Sicily Strait (Figure \ref{fig:regions_spectra} e-h), the bottom-trapped diurnal internal tide is more prominent, while there is only a small peak of semidiurnal baroclinic energy density: demonstrating further the result of Figure \ref{fig:mean_ke}, which showed that the M2 internal tide is relatively important in the western basin, while K1 is more prominent in the central Mediterranean Sea. Contrastingly, in the Ionian Sea (Figures \ref{fig:regions_spectra} i-l), energy density from the currents depends strongly on depth, with some semidiurnal power in the upper layers. As discussed in \cite{MCG23}, this region shows some interaction between the diurnal internal tide and near-inertial waves, particularly in NEMO, where the near-inertial peaks stretch toward the diurnal frequency. This analysis of the baroclinic kinetic energy further confirms that the wave-wave interactions seen in \cite{MCG23} refer to interactions between internal tides and near-inertial waves.

\subsection{Internal tide wavelengths}

Typical wavelengths of the first two modes of the M2 internal tide in the global ocean are 100-160 km and 45-80 km \cite[]{LvSM15}. Several regions are chosen for wavenumber analysis, following propagation paths of the M2 internal tide seen in Figures \ref{fig:m2_150}-\ref{fig:m2_1000}. Figure \ref{fig:domain_map} includes a map of these new analysis regions. Wavenumber spectra of the tidal baroclinic velocity of the M2 component is calculated along these paths using the periodogram as described in \ref{spectra}. The data used to calculate the periodogram is from the harmonic analysis of baroclinic currents for each tidal component, along the length of the paths of the blue highlighted boxes in Figure \ref{fig:domain_map}. Several parallel lines (five for the Algerian Sea, ten for the other regions are averaged after the periodogram is calculated for each line. 

\begin{figure}[]
\centering
\includegraphics[width=\linewidth, trim=0cm 0cm 0cm 0cm, clip]{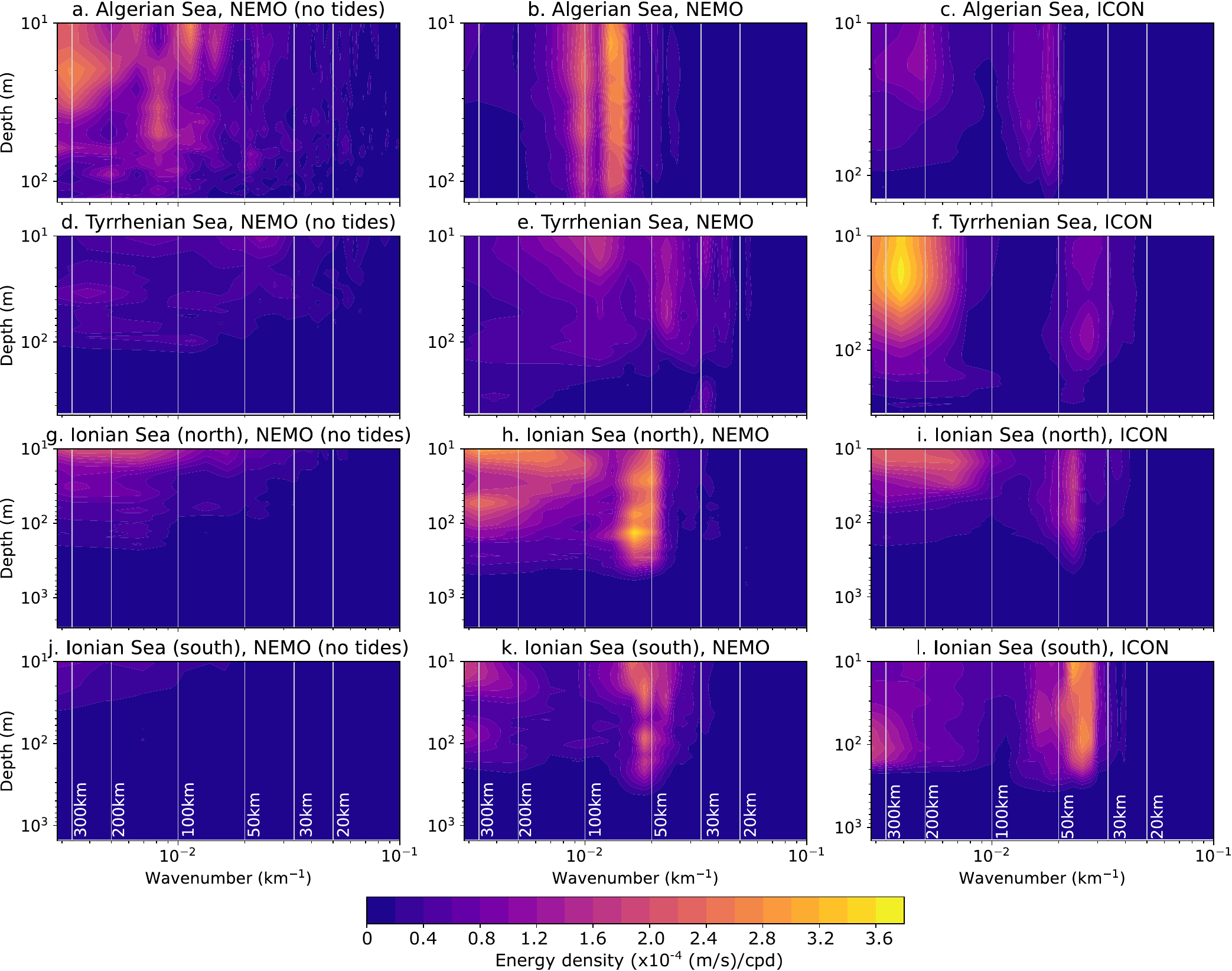}
\caption{Wavenumber energy density spectra of M2 baroclinic currents for four regions (as in Figure \ref{fig:domain_map}): a-c. Zonal velocity, Algerian Sea, NEMO (without tides), NEMO and ICON, d-f. Meridional velocity, Tyrrhenian Sea, NEMO (without tides), NEMO and ICON, g-i. Zonal velocity, Ionian Sea (north), NEMO (without tides), NEMO and ICON, j-l. Zonal velocity, Ionian Sea (south), NEMO (without tides), NEMO and ICON.}
\label{fig:m2_wavenumber}
\end{figure}

The wavenumber spectra for the M2 component along each of these paths are in Figure \ref{fig:m2_wavenumber}, with additional data from a similar NEMO configuration without tides \cite[]{MCG23}. In Figure \ref{fig:m2_wavenumber}, the largest density peak in the Tyrrhenian and northern Ionian seas are at wavelengths greater than 200km. These can be attributed to seiches in the Mediterranean Sea, which include modes at frequencies close to that of the M2 tidal component, at 11.4 hours and 12 hours \cite[]{SR83, LC95}, which are known to interact with tides \cite[]{MCG23}. The peak most relevant to internal tides is at around 50-100 km and is visible in the Algerian Sea and Ionian Sea, in both experiments but most prominently NEMO. This is the first vertical mode of the M2 internal tide. A second mode also appears in Figure \ref{fig:m2_wavenumber} b and e, at a higher wavenumber than the first, close to 50 km. 

These results from model outputs are also compared to theoretical values for the first modes of the M2 internal tide in the same region, calculated by solving the Sturm-Liouville eigenvalue problem, assuming a flat bottom within each region at its area-weighted mean depth.  Table \ref{tab:sturmliouville} shows these results for both models. In most of the regions analysed, these calculated wavelengths are well-matched with those in the spectra. It also helps us to differentiate between the peaks due to internal tides in Fig. \ref{fig:m2_wavenumber} and other peaks which are arising from other phenomena at frequencies close to that of the M2 tide, such as seiches. We argue now that the longer wavelength peaks in the Tyrrhenian and Ionian Seas are not due to tides but to seiches that are at similar frequencies. More details on the Sturm-Liouville eigenvalue problem and its associated assumptions can be found in \ref{sturmliouville}. 

\begin{table}[]
    \centering
    \begin{tabularx}{\linewidth}{|X|X|X|}
    \hline
    \textbf{Region} & \textbf{NEMO (km)} & \textbf{ICON (km)} \\
    \hline
    Algerian Sea (Fig. \ref{fig:m2_wavenumber}a-c) & 71.6 & 54.7 \\
    Tyrrhenian Sea (Fig. \ref{fig:m2_wavenumber}d-f) & 57.5 & 45.8 \\
    Ionian Sea (north) (Fig. \ref{fig:m2_wavenumber}g-i) & 52.0 & 43.3 \\
    Ionian Sea (south) (Fig. \ref{fig:m2_wavenumber}j-l) & 53.3 & 41.3 \\ 
    \hline
    \end{tabularx}
    \caption{Wavelengths of the first baroclinic mode of the M2 internal tide in the regions from Figs. \ref{fig:domain_map} and \ref{fig:m2_wavenumber}, calculated by solving the Sturm-Liouville eigenvalue problem}
    \label{tab:sturmliouville}
\end{table}

\section{Assessment of model differences}
\label{differences}

The two models broadly show internal tide generation and propagation in the same regions of the Mediterranean Sea. However, although they are similar in resolution and have some common features (see Section \ref{models}), the models show several differences in their representation of internal tides, particularly in the wavenumber analysis (see Fig. \ref{tab:sturmliouville}). There are several possible reasons for this, including the differences in the implementation of barotropic tides. An intercomparison of barotropic tides between the two models is shown in the figures in \ref{intercomparison}. Internal tide generation and propagation is also affected by the stratification and bathymetry of the models. 

It is expected that models with a more stratified ocean would be more likely to produce stronger internal tides. The energy per unit volume in a propagating internal tide at the semidiurnal frequency, $E_f$ according to \cite{B82} and \cite{GST10}, is: 

\begin{equation}
    E_f = 0.5\rho_0 (u'^2 + w'^2 + N^2 \eta^2)
\end{equation}

where $\rho_0$ is the constant reference density, $u'$ and $w'$ are perturbation velocities in the horizontal and vertical directions respectively, $N$ is the Brunt-Väisälä frequency, and $\eta$ is the displacement of a streamline of the internal wave field, $\psi$, which is itself defined by $u_i = – \psi_z$, $w_i = – \psi_x$ \cite[]{B82}. 

\begin{figure}[]
\centering
\includegraphics[width=\linewidth, trim=0cm 0cm 0cm 0cm, clip]{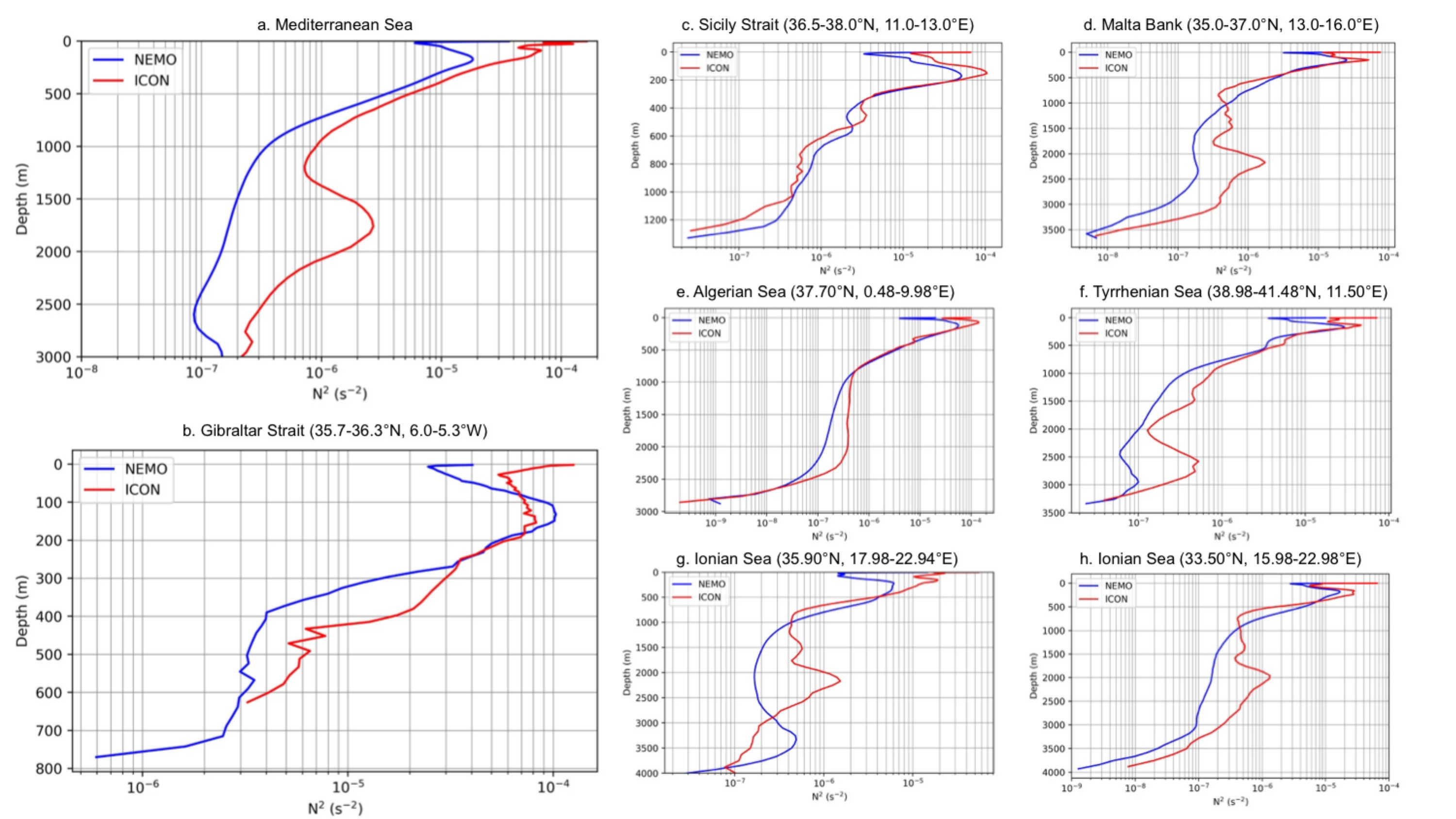}
\caption{Vertical profiles of the Brunt Väisälä frequency for the NEMO simulation (blue) and the ICON simulation (red), for a. the Mediterranean Sea, and the regions shown in Figure \ref{fig:domain_map}. These are b. the Gibraltar Strait, c. Sicily Strait, d. Malta Bank, e. Algerian Sea, f. Tyrrhenian Sea, g. Ionian Sea (north), and h. Ionian Sea (south).}
\label{fig:profiles_nemo_icon}
\end{figure}

Figure \ref{fig:profiles_nemo_icon} shows profiles of the squared Brunt-Väisälä frequency, both as a Mediterranean Sea average and for some of the key regions discussed in previous sections. We see that for the basin mean profile (Fig. \ref{fig:profiles_nemo_icon}a), the water column provided by the ICON simulation is more stratified than that of NEMO across all vertical levels. This explains the larger kinetic energy and longer wave propagation seen at many depths such as in Figure \ref{fig:m2_300} in the Eastern Mediterranean. The shape of the two profiles also differs: the profile from the ICON simulation has a second peak at around 1750m, while the NEMO experiment profile only has a subsurface peak. A Mediterranean Sea observed profile such as those in \cite{CBM12} typically looks more like that of the NEMO experiment, with only one peak close to the surface, whereas the ICON experiment’s profile has a double peak which appears more like a typical Atlantic Ocean profile such as those from \cite{ELM84}. This could lead to higher or overestimated propagation of internal tides in ICON when compared to NEMO since ICON is more stratified in several key regions for internal tide propagation. However, NEMO is the model with more generation of internal tides in the Mediterranean Sea, meaning that assessing the stratification alone doesn’t give the full picture of the model differences.  

\begin{figure}[]
\centering
\includegraphics[width=\linewidth, trim=0cm 0cm 0cm 0cm, clip]{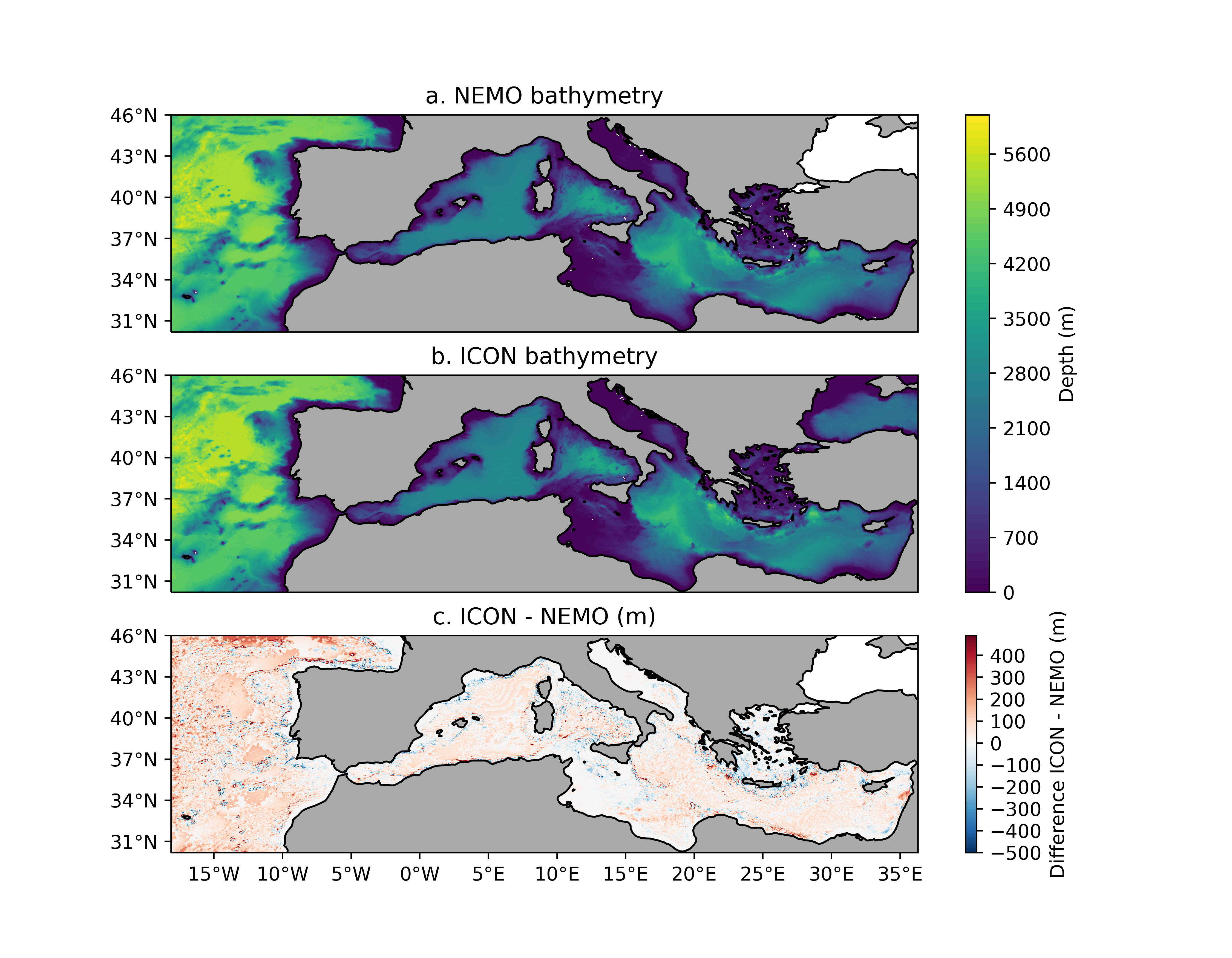}
\caption{Model bathymetry for a. NEMO, b. ICON, and c. Difference (ICON - NEMO).}
\label{fig:bathy_diff}
\end{figure}

The generation of internal tides is sensitive to changes in bathymetry since internal tides are generated along the topography. The model bathymetry, as described in Table \ref{tab:NEMO_ICON}, comes from different sources in each experiment, and although they have the same resolution (30 arc-seconds), it is interpolated differently in each model due to the differing model horizontal grids. These bathymetry variations, although they are usually small, could impact the generation regions of internal tides, and therefore their energy and direction of propagation. Figure \ref{fig:bathy_diff} shows the bathymetry differences between the two models. It is clear that although the differences in bathymetry are not large in most of the domain, the ICON bathymetry is slightly deeper than that of NEMO in many areas. However, as was shown in Figure \ref{fig:bathy_diff_regions}, we can see that this is not the case in some key regions. 

\section{Conclusions}
\label{conclusion}

For the first time, internal tides in the Mediterranean Sea were mapped in the whole basin. Three major generation sites are found: firstly, the Gibraltar Strait and Alboran Sea, secondly, the Sicily Strait and Malta Bank in the central Mediterranean Sea, and finally, the Hellenic Arc, as well as several minor generation sites in the Eastern Mediterranean more broadly, and the Tyrrhenian Sea. Both diurnal and semidiurnal internal tides were analysed, with a particular focus on the M2 and K1 components. It was found that the two components have different importance in different regions, with M2 being more prominent in the Western Mediterranean Sea and Cretan Passage, and K1 being most important in the central and southeastern parts of the basin. Semidiurnal internal tides propagate for up to hundreds of kilometres in the Mediterranean Sea, both in the eastern and western basins, particularly in the Algerian Sea, Tyrrhenian Sea, and Ionian Sea, which respectively propagate from the three key generation sites listed above. The first two modes of the M2 internal tide are resolved by the numerical models and have wavelengths comparable to \cite{LvSM15} in the NEMO experiment, whereas the ICON-O simulation has shorter wavelengths than suggested by the literature in several regions probably due to bathymetric differences.  

The two numerical experiments found similar generation sites and propagation regions for internal tides, but the specific wavelengths, the propagation direction, and propagation distance varies between the two experiments. Some reasons for this include differences in stratification, model bathymetry, and the calculation of barotropic tides, as well as particularly differing configurations of key regions such as the Gibraltar Strait between the two model implementations. The barotropic tidal amplitudes and phases provided by the NEMO and ICON experiments are compared to a global barotropic model used as a reference \cite[]{ER03} in \ref{intercomparison}. However, further validation through satellite data at the sea surface, as well as using data from cruises in some isolated regions, such as that of \cite{OPF23}, could be used to better understand which model has a more correct representation of internal tides.

This work was a first step in mapping internal tides for the first time throughout the Mediterranean basin using a high-resolution numerical model. Next steps for the study of internal tides in the Mediterranean Sea could include an analysis of interactions with other mesoscale phenomena, particularly near-inertial waves, as was briefly discussed in \cite{MCG23} as well as in this work, and the interaction between internal tides and mesoscale eddies, which are a key feature in the Mediterranean Sea circulation. It is possible that internal tides could be generated in other regions of steep topography such as small islands and seamounts, and narrow straits such as the Messina Strait, which are not properly resolved by the model implementations used in this work. A regional study of the central Mediterranean Seas with sub-kilometre scale horizontal resolution would be a particularly useful step in understanding the internal tides in this region.

\section*{Acknowledgements}{This study has been conducted including support from the University of Bologna Ph.D. programme in Future Earth, Climate Change, and Societal Challenge, and the EU Copernicus Marine Service for the Mediterranean Monitoring and Forecasting Center.}

\section*{Author contributions}{BM ran the NEMO experiment, analysed the results, and wrote the manuscript. JSvS provided the ICON dataset, and oversaw the analysis of the numerical results. EC provided supervision and support in the set-up and running of the numerical models and analysis. NP planned the study and supported the analysis of model results. All authors reviewed and edited the manuscript.} 

\pagebreak
\bibliographystyle{elsarticle-harv} 
\bibliography{references}

\begin{thebibliography}{57}
\expandafter\ifx\csname natexlab\endcsname\relax\def\natexlab#1{#1}\fi
\providecommand{\url}[1]{\texttt{#1}}
\providecommand{\href}[2]{#2}
\providecommand{\path}[1]{#1}
\providecommand{\DOIprefix}{doi:}
\providecommand{\ArXivprefix}{arXiv:}
\providecommand{\URLprefix}{URL: }
\providecommand{\Pubmedprefix}{pmid:}
\providecommand{\doi}[1]{\href{http://dx.doi.org/#1}{\path{#1}}}
\providecommand{\Pubmed}[1]{\href{pmid:#1}{\path{#1}}}
\providecommand{\bibinfo}[2]{#2}
\ifx\xfnm\relax \def\xfnm[#1]{\unskip,\space#1}\fi
\bibitem[{Alford et~al.(2012)Alford, Gregg, Zervakis and Kontoyiannis}]{AGZ12}
\bibinfo{author}{Alford, M.H.}, \bibinfo{author}{Gregg, M.C.}, \bibinfo{author}{Zervakis, V.}, \bibinfo{author}{Kontoyiannis, H.}, \bibinfo{year}{2012}.
\newblock \bibinfo{title}{{Internal wave measurements on the Cycladic Plateauof the Aegean Sea}}.
\newblock \bibinfo{journal}{J. Geophys. Res.} \bibinfo{volume}{117}, \bibinfo{pages}{C01015}.
\newblock \DOIprefix\doi{10.1029/2011JC007488}.
\bibitem[{Arbic(2022)}]{A22}
\bibinfo{author}{Arbic, B.K.}, \bibinfo{year}{2022}.
\newblock \bibinfo{title}{{Incorporating tides and internal gravity waves within global ocean general circulation models: A review}}.
\newblock \bibinfo{journal}{Progr. Oceanogr.} \bibinfo{volume}{206}.
\newblock \DOIprefix\doi{10.1016/j.pocean.2022.102824}.
\bibitem[{Baines(1982)}]{B82}
\bibinfo{author}{Baines, P.G.}, \bibinfo{year}{1982}.
\newblock \bibinfo{title}{{On internal tide generation models}}.
\newblock \bibinfo{journal}{Deep Sea Research Part A. Oceanographic Research Papers} \bibinfo{volume}{29}, \bibinfo{pages}{307--338}.
\bibitem[{Becker et~al.(2009)Becker, Sandwell, Smith, Braud, Binder, Depner, Fabre, Factor, Ingalls, Kim, Ladner, Marks, Nelson, Pharaoh, Trimmer, Rosenberg, Wallace and Weatherall}]{SRTM30_bathy}
\bibinfo{author}{Becker, J.J.}, \bibinfo{author}{Sandwell, D.T.}, \bibinfo{author}{Smith, W.H.F.}, \bibinfo{author}{Braud, J.}, \bibinfo{author}{Binder, B.}, \bibinfo{author}{Depner, J.}, \bibinfo{author}{Fabre, D.}, \bibinfo{author}{Factor, J.}, \bibinfo{author}{Ingalls, S.}, \bibinfo{author}{Kim, S.H.}, \bibinfo{author}{Ladner, R.}, \bibinfo{author}{Marks, K.}, \bibinfo{author}{Nelson, S.}, \bibinfo{author}{Pharaoh, A.}, \bibinfo{author}{Trimmer, R.}, \bibinfo{author}{Rosenberg, J.V.}, \bibinfo{author}{Wallace, G.}, \bibinfo{author}{Weatherall, P.}, \bibinfo{year}{2009}.
\newblock \bibinfo{title}{{Global Bathymetry and Elevation Data at 30 Arc Seconds Resolution: SRTM30\_PLUS}}.
\newblock \bibinfo{journal}{Marine Geodesy} \bibinfo{volume}{32}, \bibinfo{pages}{355--371}.
\newblock \URLprefix \url{http://topex.ucsd.edu/sandwell/publications/124\_MG\_Becker.pdf}.
\bibitem[{Boyer et~al.(2013)Boyer, Antonov, Baranova, Coleman, Garcia, Grodsky, Johnson, Locarnini, Mishonov, O'Brien, Paver, Reagan, Seidov, Smolyar,  and Zweng}]{WORLD_OCEAN_ATLAS}
\bibinfo{author}{Boyer, T.}, \bibinfo{author}{Antonov, J.I.}, \bibinfo{author}{Baranova, O.K.}, \bibinfo{author}{Coleman, C.}, \bibinfo{author}{Garcia, H.E.}, \bibinfo{author}{Grodsky, A.}, \bibinfo{author}{Johnson, D.R.}, \bibinfo{author}{Locarnini, R.A.}, \bibinfo{author}{Mishonov, A.V.}, \bibinfo{author}{O'Brien, T.}, \bibinfo{author}{Paver, C.}, \bibinfo{author}{Reagan, J.}, \bibinfo{author}{Seidov, D.}, \bibinfo{author}{Smolyar, I.V.}, , \bibinfo{author}{Zweng, M.M.}, \bibinfo{year}{2013}.
\newblock \bibinfo{title}{{World Ocean Database 2013, NOAA Atlas NESDIS 72}}. \bibinfo{address}{Silver Spring, MD}.
\newblock \DOIprefix\doi{10.7289/V5NZ85MT}.
\bibitem[{Buijsman et~al.(2014)Buijsman, Klymak, Legg, Alford, Farmer, MacKinnon, Nash, Park, Pickering and Simmons}]{BKL14}
\bibinfo{author}{Buijsman, M.C.}, \bibinfo{author}{Klymak, J.M.}, \bibinfo{author}{Legg, S.}, \bibinfo{author}{Alford, M.H.}, \bibinfo{author}{Farmer, D.}, \bibinfo{author}{MacKinnon, J.A.}, \bibinfo{author}{Nash, J.D.}, \bibinfo{author}{Park, J.H.}, \bibinfo{author}{Pickering, A.}, \bibinfo{author}{Simmons, H.}, \bibinfo{year}{2014}.
\newblock \bibinfo{title}{{Three-Dimensional Double-Ridge Internal Tide Resonance in Luzon Strait}}.
\newblock \bibinfo{journal}{J. Phys. Oceanogr.} \bibinfo{volume}{44}, \bibinfo{pages}{850 -- 869}.
\newblock \DOIprefix\doi{10.1175/JPO-D-13-024.1}.
\bibitem[{Candela et~al.(1990)Candela, Winant and Ruiz}]{CWR90}
\bibinfo{author}{Candela, J.}, \bibinfo{author}{Winant, C.}, \bibinfo{author}{Ruiz, A.}, \bibinfo{year}{1990}.
\newblock \bibinfo{title}{{Tides in the Strait of Gibraltar}}.
\newblock \bibinfo{journal}{J. Geophys. Res. Oceans} \bibinfo{volume}{95}, \bibinfo{pages}{7313--7335}.
\newblock \DOIprefix\doi{10.1029/JC095iC05p07313}.
\bibitem[{Clementi et~al.(2021)Clementi, Aydogdu, Goglio, Pistoia, Escudier, Drudi, Grandi, Mariani, Lyubartsev, Lecci, Cretí, Coppini, Masina and Pinardi}]{CMEMS_EAS6}
\bibinfo{author}{Clementi, E.}, \bibinfo{author}{Aydogdu, A.}, \bibinfo{author}{Goglio, A.C.}, \bibinfo{author}{Pistoia, J.}, \bibinfo{author}{Escudier, R.}, \bibinfo{author}{Drudi, M.}, \bibinfo{author}{Grandi, A.}, \bibinfo{author}{Mariani, A.}, \bibinfo{author}{Lyubartsev, V.}, \bibinfo{author}{Lecci, R.}, \bibinfo{author}{Cretí, S.}, \bibinfo{author}{Coppini, G.}, \bibinfo{author}{Masina, S.}, \bibinfo{author}{Pinardi, N.}, \bibinfo{year}{2021}.
\newblock \bibinfo{title}{{Mediterranean Sea Physical Analysis and Forecast (CMEMS MED-Currents, EAS6 system)}}.
\newblock \bibinfo{journal}{Copernicus Monitoring Environment Marine Service (CMEMS)} \DOIprefix\doi{10.25423/CMCC/MEDSEA\_ANALYSISFORECAST\_PHY\_006\_013\_EAS6}.
\bibitem[{Coppini et~al.(2023)Coppini, Clementi, Cossarini, Salon, Korres, Ravdas, Lecci, Pistoia, Goglio, Drudi, Grandi, Aydogdu, Escudier, Cipollone, Lyubartsev, Mariani, Cret\`{\i}, Palermo, Scuro, Masina, Pinardi, Navarra, Delrosso, Teruzzi, Di~Biagio, Bolzon, Feudale, Coidessa, Amadio, Brosich, Mir\'o, Alvarez, Lazzari, Solidoro, Oikonomou and Zacharioudaki}]{CCC23}
\bibinfo{author}{Coppini, G.}, \bibinfo{author}{Clementi, E.}, \bibinfo{author}{Cossarini, G.}, \bibinfo{author}{Salon, S.}, \bibinfo{author}{Korres, G.}, \bibinfo{author}{Ravdas, M.}, \bibinfo{author}{Lecci, R.}, \bibinfo{author}{Pistoia, J.}, \bibinfo{author}{Goglio, A.C.}, \bibinfo{author}{Drudi, M.}, \bibinfo{author}{Grandi, A.}, \bibinfo{author}{Aydogdu, A.}, \bibinfo{author}{Escudier, R.}, \bibinfo{author}{Cipollone, A.}, \bibinfo{author}{Lyubartsev, V.}, \bibinfo{author}{Mariani, A.}, \bibinfo{author}{Cret\`{\i}, S.}, \bibinfo{author}{Palermo, F.}, \bibinfo{author}{Scuro, M.}, \bibinfo{author}{Masina, S.}, \bibinfo{author}{Pinardi, N.}, \bibinfo{author}{Navarra, A.}, \bibinfo{author}{Delrosso, D.}, \bibinfo{author}{Teruzzi, A.}, \bibinfo{author}{Di~Biagio, V.}, \bibinfo{author}{Bolzon, G.}, \bibinfo{author}{Feudale, L.}, \bibinfo{author}{Coidessa, G.}, \bibinfo{author}{Amadio, C.}, \bibinfo{author}{Brosich, A.}, \bibinfo{author}{Mir\'o, A.}, \bibinfo{author}{Alvarez, E.}, \bibinfo{author}{Lazzari, P.},
  \bibinfo{author}{Solidoro, C.}, \bibinfo{author}{Oikonomou, C.}, \bibinfo{author}{Zacharioudaki, A.}, \bibinfo{year}{2023}.
\newblock \bibinfo{title}{{The Mediterranean forecasting system. Part I: evolution and performance}}.
\newblock \bibinfo{journal}{EGUsphere} , \bibinfo{pages}{1--50}\DOIprefix\doi{10.5194/egusphere-2022-1337}.
\bibitem[{Cuypers et~al.(2012)Cuypers, Bouruet-Aubertot, Marec and Fuda}]{CBM12}
\bibinfo{author}{Cuypers, Y.}, \bibinfo{author}{Bouruet-Aubertot, P.}, \bibinfo{author}{Marec, C.}, \bibinfo{author}{Fuda, J.L.}, \bibinfo{year}{2012}.
\newblock \bibinfo{title}{{Characterization of turbulence from a fine-scale parameterization and microstructure measurements in the Mediterranean Sea during the BOUM experiment}}.
\newblock \bibinfo{journal}{Biogeosciences} \bibinfo{volume}{9}, \bibinfo{pages}{3131--3149}.
\newblock \DOIprefix\doi{10.5194/bg-9-3131-2012}.
\bibitem[{{Deliverable of Perseus}(2012)}]{P12}
\bibinfo{author}{{Deliverable of Perseus}}, \bibinfo{year}{2012}.
\newblock \bibinfo{title}{{Deliverable D4.6, SES land-based runoff and nutrient load data (1980–2000), edited by: Bouwman, L. and van Apeldoorn, D.}}
\newblock \bibinfo{journal}{2012 PERSEUS H2020 grant agreement n. 287600, European Commission} \URLprefix \url{http://www.perseus-net. eu/assets/media/PDF/deliverables/3321 6\_Final.pdf}.
\bibitem[{Demiraj et~al.(1996)Demiraj, Bicja, Gjika, Gjiknuri, Muçaj, Hoxha, Hoxha, Karadumi, Kongoli, Mullaj, Mustaqi, Palluqi, Ruli, Selfo, Shehi and Sino}]{D96}
\bibinfo{author}{Demiraj, E.}, \bibinfo{author}{Bicja, M.}, \bibinfo{author}{Gjika, E.}, \bibinfo{author}{Gjiknuri, L.}, \bibinfo{author}{Muçaj, L.G.}, \bibinfo{author}{Hoxha, F.}, \bibinfo{author}{Hoxha, P.}, \bibinfo{author}{Karadumi, S.}, \bibinfo{author}{Kongoli, S.}, \bibinfo{author}{Mullaj, A.}, \bibinfo{author}{Mustaqi, V.}, \bibinfo{author}{Palluqi, A.}, \bibinfo{author}{Ruli, E.}, \bibinfo{author}{Selfo, M.}, \bibinfo{author}{Shehi, A.}, \bibinfo{author}{Sino, Q.}, \bibinfo{year}{1996}.
\newblock \bibinfo{title}{{Implications of climate change for the Albanian Coast, Mediterranean Action Plan, MAP Technical Reports Series}}.
\newblock \bibinfo{number}{98}, \bibinfo{publisher}{UNEP}.
\bibitem[{Dunphy and Lamb(2014)}]{DL14}
\bibinfo{author}{Dunphy, M.}, \bibinfo{author}{Lamb, K.G.}, \bibinfo{year}{2014}.
\newblock \bibinfo{title}{{Focusing and vertical mode scattering of the first mode internal tide by mesoscale eddy interaction}}.
\newblock \bibinfo{journal}{J. Geophys. Res. Oceans} \bibinfo{volume}{119}, \bibinfo{pages}{523--536}.
\newblock \DOIprefix\doi{10.1002/2013JC009293}.
\bibitem[{Egbert and Erofeeva(2002)}]{EE02}
\bibinfo{author}{Egbert, G.D.}, \bibinfo{author}{Erofeeva, S.Y.}, \bibinfo{year}{2002}.
\newblock \bibinfo{title}{{Efficient inverse modeling of barotropic ocean tides}}.
\newblock \bibinfo{journal}{J. Atmosph. Oceanic Tech.} \bibinfo{volume}{19.2}, \bibinfo{pages}{183--204}.
\bibitem[{Egbert and Ray(2003)}]{ER03}
\bibinfo{author}{Egbert, G.D.}, \bibinfo{author}{Ray, R.D.}, \bibinfo{year}{2003}.
\newblock \bibinfo{title}{{Semi-diurnal and diurnal tidal dissipation from TOPEX/Poseidon altimetry}}.
\newblock \bibinfo{journal}{Geophys. Res. Lett.} \bibinfo{volume}{30}, \bibinfo{pages}{1907}.
\newblock \DOIprefix\doi{10.1029/2003GL017676}.
\bibitem[{Emery et~al.(1984)Emery, Lee and Magaard}]{ELM84}
\bibinfo{author}{Emery, W.J.}, \bibinfo{author}{Lee, W.G.}, \bibinfo{author}{Magaard, L.}, \bibinfo{year}{1984}.
\newblock \bibinfo{title}{{Geographic and Seasonal Distributions of Brunt–Väisälä Frequency and Rossby Radii in the North Pacific and North Atlantic}}.
\newblock \bibinfo{journal}{J. Phys. Oceanogr.} \bibinfo{volume}{14}, \bibinfo{pages}{294--317}.
\newblock \DOIprefix\doi{10.1175/1520-0485(1984)014<0294:GASDOB>2.0.CO;2}.
\bibitem[{Fakete et~al.(1999)Fakete, Vörösmarty and Grabs}]{FVG99}
\bibinfo{author}{Fakete, B.}, \bibinfo{author}{Vörösmarty, C.}, \bibinfo{author}{Grabs, W.}, \bibinfo{year}{1999}.
\newblock \bibinfo{title}{{Global composite runoff fields based on observed river discharge and simulated water balances}}.
\newblock \bibinfo{journal}{Technical Report 22, Global Runoff Data Centre, Koblenz, Germany} .
\bibitem[{Foreman et~al.(2009)Foreman, Cherniawsky and Ballantyne}]{FCB09}
\bibinfo{author}{Foreman, M.G.G.}, \bibinfo{author}{Cherniawsky, J.Y.}, \bibinfo{author}{Ballantyne, V.A.}, \bibinfo{year}{2009}.
\newblock \bibinfo{title}{{Versatile Harmonic Tidal Analysis: Improvements and Applications}}.
\newblock \bibinfo{journal}{J. Atmos. Oceanic Technol.} \bibinfo{volume}{26}, \bibinfo{pages}{806–817}.
\newblock \DOIprefix\doi{10.1175/2008JTECHO615.1}.
\bibitem[{Galloudec et~al.(2022)Galloudec, Chune, Nouel, Fernandez, Derval, Tressol, Dussurget, Biardeau and Tonani}]{CMEMS_global}
\bibinfo{author}{Galloudec, O.L.}, \bibinfo{author}{Chune, S.L.}, \bibinfo{author}{Nouel, L.}, \bibinfo{author}{Fernandez, E.}, \bibinfo{author}{Derval, C.}, \bibinfo{author}{Tressol, M.}, \bibinfo{author}{Dussurget, R.}, \bibinfo{author}{Biardeau, A.}, \bibinfo{author}{Tonani, M.}, \bibinfo{year}{2022}.
\newblock \bibinfo{title}{{Global Ocean Physical Analysis and Forecasting Product}}.
\newblock \bibinfo{journal}{Copernicus Monitoring Environment Marine Service (CMEMS)} \DOIprefix\doi{10.48670/moi-00016}.
\bibitem[{Gaspar et~al.(1990)Gaspar, Grégoris and Lefevre}]{GGL90}
\bibinfo{author}{Gaspar, P.}, \bibinfo{author}{Grégoris, Y.}, \bibinfo{author}{Lefevre, J.M.}, \bibinfo{year}{1990}.
\newblock \bibinfo{title}{{A simple eddy kinetic energy model for simulations of the oceanic vertical mixing: Tests at station Papa and long-term upper ocean study site}}.
\newblock \bibinfo{journal}{J. Geophys. Res. Oceans} \bibinfo{volume}{95}, \bibinfo{pages}{16179--16193}.
\newblock \DOIprefix\doi{10.1029/JC095iC09p16179}.
\bibitem[{Gasparini et~al.(2004)Gasparini, Smeed, Alderson, Sparnocchia, Vetrano and Mazzola}]{GSA04}
\bibinfo{author}{Gasparini, G.P.}, \bibinfo{author}{Smeed, D.A.}, \bibinfo{author}{Alderson, S.}, \bibinfo{author}{Sparnocchia, S.}, \bibinfo{author}{Vetrano, A.}, \bibinfo{author}{Mazzola, S.}, \bibinfo{year}{2004}.
\newblock \bibinfo{title}{{Tidal and subtidal currents in the Strait of Sicily}}.
\newblock \bibinfo{journal}{J. Geophys. Res.} \bibinfo{volume}{109}, \bibinfo{pages}{C02011}.
\newblock \DOIprefix\doi{10.1029/2003JC002011}.
\bibitem[{Gates et~al.(1993)Gates, Hagemann and Golz}]{GHG93}
\bibinfo{author}{Gates, L.D.}, \bibinfo{author}{Hagemann, S.}, \bibinfo{author}{Golz, C.}, \bibinfo{year}{1993}.
\newblock \bibinfo{title}{{Observed historical discharge data from major rivers for climate model validation}}.
\newblock \bibinfo{journal}{Tech. Rep.} \bibinfo{volume}{307}.
\bibitem[{{GEBCO Bathymetric Compilation Group 2014}(2014)}]{GEBCO}
\bibinfo{author}{{GEBCO Bathymetric Compilation Group 2014}}, \bibinfo{year}{2014}.
\newblock \bibinfo{title}{{The GEBCO\_2014 Grid, version 20150318}}.
\newblock \bibinfo{journal}{{NERC EDS British Oceanographic Data Centre NOC}} \URLprefix \url{www.gebco.net}.
\bibitem[{Giorgetta et~al.(2018)Giorgetta, Brokopf, Crueger, Esch, Fiedler, Helmert, Hohenegger, Kornblueh, Köhler, Manzini, Mauritsen, Nam, Raddatz, Rast, Reinert, Sakradzija, Schmidt, Schneck, Schnur, Silvers, Wan, Zängl and Stevens}]{GBC18}
\bibinfo{author}{Giorgetta, M.A.}, \bibinfo{author}{Brokopf, R.}, \bibinfo{author}{Crueger, T.}, \bibinfo{author}{Esch, M.}, \bibinfo{author}{Fiedler, S.}, \bibinfo{author}{Helmert, J.}, \bibinfo{author}{Hohenegger, C.}, \bibinfo{author}{Kornblueh, L.}, \bibinfo{author}{Köhler, M.}, \bibinfo{author}{Manzini, E.}, \bibinfo{author}{Mauritsen, T.}, \bibinfo{author}{Nam, C.}, \bibinfo{author}{Raddatz, T.}, \bibinfo{author}{Rast, S.}, \bibinfo{author}{Reinert, D.}, \bibinfo{author}{Sakradzija, M.}, \bibinfo{author}{Schmidt, H.}, \bibinfo{author}{Schneck, R.}, \bibinfo{author}{Schnur, R.}, \bibinfo{author}{Silvers, L.}, \bibinfo{author}{Wan, H.}, \bibinfo{author}{Zängl, G.}, \bibinfo{author}{Stevens, B.}, \bibinfo{year}{2018}.
\newblock \bibinfo{title}{{ICON-A, the Atmosphere Component of the ICON Earth System Model: I. Model Description}}.
\newblock \bibinfo{journal}{J. Adv. Modell. Earth Systems} \bibinfo{volume}{10}, \bibinfo{pages}{1613--1637}.
\newblock \DOIprefix\doi{10.1029/2017MS001242}.
\bibitem[{Green et~al.(2010)Green, Simpson, Thorpe and Rippeth}]{GST10}
\bibinfo{author}{Green, J.}, \bibinfo{author}{Simpson, J.}, \bibinfo{author}{Thorpe, S.}, \bibinfo{author}{Rippeth, T.}, \bibinfo{year}{2010}.
\newblock \bibinfo{title}{{Observations of internal tidal waves in the isolated seasonally stratified region of the western Irish Sea}}.
\newblock \bibinfo{journal}{Continental Shelf Res.} \bibinfo{volume}{30}, \bibinfo{pages}{214--225}.
\newblock \DOIprefix\doi{10.1016/j.csr.2009.11.004}.
\bibitem[{Guo et~al.(2023)Guo, Wang, Cao, Xie, Song and Guo}]{GWC23}
\bibinfo{author}{Guo, Z.}, \bibinfo{author}{Wang, S.}, \bibinfo{author}{Cao, A.}, \bibinfo{author}{Xie, J.}, \bibinfo{author}{Song, J.}, \bibinfo{author}{Guo, X.}, \bibinfo{year}{2023}.
\newblock \bibinfo{title}{{Refraction of the M2 internal tides by mesoscale eddies in the South China Sea}}.
\newblock \bibinfo{journal}{Deep Sea Res. Part I: Oceanogr. Res. Pap.} \bibinfo{volume}{192}, \bibinfo{pages}{103946}.
\newblock \DOIprefix\doi{10.1016/j.dsr.2022.103946}.
\bibitem[{Hersbach et~al.(2023)Hersbach, Bell, Berrisford, Biavati, Horányi, Muñoz~Sabater, Nicolas, Peubey, Radu, Rozum, Schepers, Simmons, Soci, Dee and Thépaut}]{ERA5}
\bibinfo{author}{Hersbach, H.}, \bibinfo{author}{Bell, B.}, \bibinfo{author}{Berrisford, P.}, \bibinfo{author}{Biavati, G.}, \bibinfo{author}{Horányi, A.}, \bibinfo{author}{Muñoz~Sabater, J.}, \bibinfo{author}{Nicolas, J.}, \bibinfo{author}{Peubey, C.}, \bibinfo{author}{Radu, R.}, \bibinfo{author}{Rozum, I.}, \bibinfo{author}{Schepers, D.}, \bibinfo{author}{Simmons, A.}, \bibinfo{author}{Soci, C.}, \bibinfo{author}{Dee, D.}, \bibinfo{author}{Thépaut, J.N.}, \bibinfo{year}{2023}.
\newblock \bibinfo{title}{{ERA5 hourly data on single levels from 1979 to present (Copernicus Climate Change Service (C3S) Climate Data Store (CDS))}}.
\newblock \bibinfo{journal}{{Copernicus Climate Change Service (C3S) Climate Data Store (CDS)}} \DOIprefix\doi{10.24381/cds.adbb2d47}.
\bibitem[{Kang and Fringer(2012)}]{KF12}
\bibinfo{author}{Kang, D.}, \bibinfo{author}{Fringer, O.}, \bibinfo{year}{2012}.
\newblock \bibinfo{title}{{Energetics of Barotropic and Baroclinic Tides in the Monterey Bay Area}}.
\newblock \bibinfo{journal}{J. Phys. Oceanogr.} \bibinfo{volume}{42}, \bibinfo{pages}{272 -- 290}.
\newblock \DOIprefix\doi{10.1175/JPO-D-11-039.1}.
\bibitem[{Kelly and Lermusiaux(2016)}]{KL16}
\bibinfo{author}{Kelly, S.M.}, \bibinfo{author}{Lermusiaux, P.F.J.}, \bibinfo{year}{2016}.
\newblock \bibinfo{title}{{Internal-tide interactions with the Gulf Stream and Middle Atlantic Bight shelfbreak front}}.
\newblock \bibinfo{journal}{J. Geophys. Res. Oceans} \bibinfo{volume}{121}, \bibinfo{pages}{6271–6294}.
\newblock \DOIprefix\doi{10.1002/2016JC011639}.
\bibitem[{Korn et~al.(2022)Korn, Brüggemann, Jungclaus, Lorenz, Gutjahr, Haak, Linardakis, Mehlmann, Mikolajewicz, Notz, Putrasahan, Singh, von Storch, Zhu and Marotzke}]{KBJ22}
\bibinfo{author}{Korn, P.}, \bibinfo{author}{Brüggemann, N.}, \bibinfo{author}{Jungclaus, J.H.}, \bibinfo{author}{Lorenz, S.J.}, \bibinfo{author}{Gutjahr, O.}, \bibinfo{author}{Haak, H.}, \bibinfo{author}{Linardakis, L.}, \bibinfo{author}{Mehlmann, C.}, \bibinfo{author}{Mikolajewicz, U.}, \bibinfo{author}{Notz, D.}, \bibinfo{author}{Putrasahan, D.A.}, \bibinfo{author}{Singh, V.}, \bibinfo{author}{von Storch, J.S.}, \bibinfo{author}{Zhu, X.}, \bibinfo{author}{Marotzke, J.}, \bibinfo{year}{2022}.
\newblock \bibinfo{title}{{ICON-O: The Ocean Component of the ICON Earth System Model—Global Simulation Characteristics and Local Telescoping Capability}}.
\newblock \bibinfo{journal}{J. Adv. Modell. Earth Syst.} \bibinfo{volume}{14}, \bibinfo{pages}{e2021MS002952}.
\newblock \DOIprefix\doi{10.1029/2021MS002952}.
\bibitem[{Lafuente et~al.(2000)Lafuente, Vargas, Plaza, Sarhan, Candela and Bascheck}]{LVP00}
\bibinfo{author}{Lafuente, J.G.}, \bibinfo{author}{Vargas, J.M.}, \bibinfo{author}{Plaza, F.}, \bibinfo{author}{Sarhan, T.}, \bibinfo{author}{Candela, J.}, \bibinfo{author}{Bascheck, B.}, \bibinfo{year}{2000}.
\newblock \bibinfo{title}{{Tide at the eastern section of the Strait of Gibraltar}}.
\newblock \bibinfo{journal}{J. Geophys. Res. Oceans} \bibinfo{volume}{105}, \bibinfo{pages}{14197– 14213}.
\newblock \DOIprefix\doi{10.1029/2000JC900007}.
\bibitem[{Li and von Storch(2020)}]{LvS20}
\bibinfo{author}{Li, Z.}, \bibinfo{author}{von Storch, J.S.}, \bibinfo{year}{2020}.
\newblock \bibinfo{title}{{M2 Internal-Tide Generation in STORMTIDE2}}.
\newblock \bibinfo{journal}{J. Geophys. Res. Oceans} \bibinfo{volume}{125}, \bibinfo{pages}{e2019JC015453}.
\newblock \DOIprefix\doi{https://doi.org/10.1029/2019JC015453}.
\bibitem[{Li et~al.(2015)Li, {von Storch} and Müller}]{LvSM15}
\bibinfo{author}{Li, Z.}, \bibinfo{author}{{von Storch}, J.S.}, \bibinfo{author}{Müller, M.}, \bibinfo{year}{2015}.
\newblock \bibinfo{title}{{The M2 Internal Tide Simulated by a 1/10° OGCM}}.
\newblock \bibinfo{journal}{J. Phys. Oceanogr.} \bibinfo{volume}{45}, \bibinfo{pages}{3119 -- 3135}.
\newblock \DOIprefix\doi{10.1175/JPO-D-14-0228.1}.
\bibitem[{Li et~al.(2017)Li, {von Storch} and Müller}]{LvSM17}
\bibinfo{author}{Li, Z.}, \bibinfo{author}{{von Storch}, J.S.}, \bibinfo{author}{Müller, M.}, \bibinfo{year}{2017}.
\newblock \bibinfo{title}{{The K1 internal tide simulated by a 1/10° OGCM}}.
\newblock \bibinfo{journal}{Ocean Modell.} \bibinfo{volume}{113}, \bibinfo{pages}{145--156}.
\newblock \DOIprefix\doi{10.1016/j.ocemod.2017.04.002}.
\bibitem[{Logemann et~al.(2021)Logemann, Linardakis, Korn and Schrum}]{LLK21}
\bibinfo{author}{Logemann, K.}, \bibinfo{author}{Linardakis, L.}, \bibinfo{author}{Korn, P.}, \bibinfo{author}{Schrum, C.}, \bibinfo{year}{2021}.
\newblock \bibinfo{title}{{Global tide simulations with ICON-O: testing the model performance on highly irregular meshes}}.
\newblock \bibinfo{journal}{Ocean Dyn.} \bibinfo{volume}{71}, \bibinfo{pages}{43--57}.
\newblock \DOIprefix\doi{10.1007/s10236-020-01428-7}.
\bibitem[{Lozano and Candela(1995)}]{LC95}
\bibinfo{author}{Lozano, C.J.}, \bibinfo{author}{Candela, J.}, \bibinfo{year}{1995}.
\newblock \bibinfo{title}{{The M2 tide in the Mediterranean Sea: Dynamic analysis and data assimilation}}.
\newblock \bibinfo{journal}{Oceanol. Acta} \bibinfo{volume}{18}, \bibinfo{pages}{419--441}.
\bibitem[{Madec et~al.(1998)Madec, Delecluse, Imbard and Levy}]{MDI98}
\bibinfo{author}{Madec, G.}, \bibinfo{author}{Delecluse, P.}, \bibinfo{author}{Imbard, M.}, \bibinfo{author}{Levy, C.}, \bibinfo{year}{1998}.
\newblock \bibinfo{title}{{OPA 8.1 Ocean general circulation model reference manual}}.
\newblock \bibinfo{journal}{Note du P{\^o}le de mod{\'e}lisation} \bibinfo{volume}{11}.
\newblock \URLprefix \url{https://www.nemo-ocean.eu/wp-content/uploads/Doc\_OPA8.1.pdf}.
\bibitem[{Maderich et~al.(2015)Maderich, Ilyin and Lemeshko}]{MIL15}
\bibinfo{author}{Maderich, V.}, \bibinfo{author}{Ilyin, Y.}, \bibinfo{author}{Lemeshko, E.}, \bibinfo{year}{2015}.
\newblock \bibinfo{title}{{Seasonal and interannual variability of the water exchange in the Turkish Straits System estimated by modelling}}.
\newblock \bibinfo{journal}{Mediterr. Mar. Sci.} \bibinfo{volume}{16}, \bibinfo{pages}{444–459}.
\newblock \DOIprefix\doi{10.12681/mms.1103}.
\bibitem[{McDonagh(2024)}]{M24}
\bibinfo{author}{McDonagh, B.}, \bibinfo{year}{2024}.
\newblock \bibinfo{title}{{Analysis of the effects of barotropic and internal tides on the Mediterranean Sea dynamics through numerical experiments}}.
\newblock Ph.D. thesis. University of Bologna.
\bibitem[{McDonagh et~al.(2024)McDonagh, Clementi, Goglio and Pinardi}]{MCG23}
\bibinfo{author}{McDonagh, B.}, \bibinfo{author}{Clementi, E.}, \bibinfo{author}{Goglio, A.C.}, \bibinfo{author}{Pinardi, N.}, \bibinfo{year}{2024}.
\newblock \bibinfo{title}{{The characteristics of tides and their effects on the general circulation of the Mediterranean Sea}}.
\newblock \bibinfo{journal}{Ocean Sci.} \bibinfo{volume}{20}, \bibinfo{pages}{1051–1066}.
\newblock \DOIprefix\doi{10.5194/os-20-1051-2024}.
\bibitem[{Merrifield et~al.(2001)Merrifield, Holloway and Johnston}]{MHJ01}
\bibinfo{author}{Merrifield, M.A.}, \bibinfo{author}{Holloway, P.E.}, \bibinfo{author}{Johnston, T.M.S.}, \bibinfo{year}{2001}.
\newblock \bibinfo{title}{{The generation of internal tides at the Hawaiian Ridge}}.
\newblock \bibinfo{journal}{Geophys. Res. Lett.} \bibinfo{volume}{28}, \bibinfo{pages}{559--562}.
\newblock \DOIprefix\doi{10.1029/2000GL011749}.
\bibitem[{Mihanović et~al.(2009)Mihanović, Orlić and Pasarić}]{MOP09}
\bibinfo{author}{Mihanović, H.}, \bibinfo{author}{Orlić, M.}, \bibinfo{author}{Pasarić, Z.}, \bibinfo{year}{2009}.
\newblock \bibinfo{title}{{Diurnal thermocline oscillations driven by tidal flow around an island in the Middle Adriatic}}.
\newblock \bibinfo{journal}{J. Marine Sys.} \bibinfo{volume}{78}, \bibinfo{pages}{S157–S168}.
\newblock \DOIprefix\doi{10.1016/j.jmarsys.2009.01.021}.
\bibitem[{Morozov et~al.(2002)Morozov, Trulsen, Velarde and Vlasenko}]{MTV02}
\bibinfo{author}{Morozov, E.G.}, \bibinfo{author}{Trulsen, K.}, \bibinfo{author}{Velarde, M.G.}, \bibinfo{author}{Vlasenko, V.I.}, \bibinfo{year}{2002}.
\newblock \bibinfo{title}{{Internal Tides in the Strait of Gibraltar}}.
\newblock \bibinfo{journal}{J. Phys. Oceanogr.} \bibinfo{volume}{32}, \bibinfo{pages}{3193--3206}.
\newblock \DOIprefix\doi{10.1175/1520-0485(2002)032<3193:ITITSO>2.0.CO;2}.
\bibitem[{Munk and Wunsch(1998)}]{MW98}
\bibinfo{author}{Munk, W.}, \bibinfo{author}{Wunsch, C.}, \bibinfo{year}{1998}.
\newblock \bibinfo{title}{{Abyssal recipes II: energetics of tidal and wind mixing}}.
\newblock \bibinfo{journal}{Deep Sea Res., Part I} \bibinfo{volume}{45}, \bibinfo{pages}{1977--2010}.
\newblock \DOIprefix\doi{10.1016/S0967-0637(98)00070-3}.
\bibitem[{Müller(2013)}]{M13}
\bibinfo{author}{Müller, M.}, \bibinfo{year}{2013}.
\newblock \bibinfo{title}{{On the space- and time-dependence of barotropic-to-baroclinic tidal energy conversion}}.
\newblock \bibinfo{journal}{Ocean Modell.} \bibinfo{volume}{72}, \bibinfo{pages}{242--252}.
\newblock \DOIprefix\doi{10.1016/j.ocemod.2013.09.007}.
\bibitem[{Müller et~al.(2012)Müller, Cherniawsky, Foreman and von Storch}]{MCF12}
\bibinfo{author}{Müller, M.}, \bibinfo{author}{Cherniawsky, J.Y.}, \bibinfo{author}{Foreman, M.G.G.}, \bibinfo{author}{von Storch, J.S.}, \bibinfo{year}{2012}.
\newblock \bibinfo{title}{{Global M2 internal tide and its seasonal variability from high resolution ocean circulation and tide modeling}}.
\newblock \bibinfo{journal}{Geophys. Res. Lett.} \bibinfo{volume}{39}.
\newblock \DOIprefix\doi{10.1029/2012GL053320}.
\bibitem[{Niwa and Hibiya(2001)}]{NH01}
\bibinfo{author}{Niwa, Y.}, \bibinfo{author}{Hibiya, T.}, \bibinfo{year}{2001}.
\newblock \bibinfo{title}{{Numerical study of the spatial distribution of the M2 internal tide in the Pacific Ocean}}.
\newblock \bibinfo{journal}{J. Geophys. Res. Oceans} \bibinfo{volume}{106}, \bibinfo{pages}{22441--22449}.
\newblock \DOIprefix\doi{10.1029/2000JC000770}.
\bibitem[{Niwa and Hibiya(2014)}]{NH14}
\bibinfo{author}{Niwa, Y.}, \bibinfo{author}{Hibiya, T.}, \bibinfo{year}{2014}.
\newblock \bibinfo{title}{Generation of baroclinic tide energy in a global three-dimensional numerical model with different spatial grid resolutions}.
\newblock \bibinfo{journal}{Ocean Modell.} \bibinfo{volume}{80}, \bibinfo{pages}{59--73}.
\newblock \DOIprefix\doi{10.1016/j.ocemod.2014.05.003}.
\bibitem[{Oddo et~al.(2023)Oddo, Poulain, Falchetti, Storto and Zappa}]{OPF23}
\bibinfo{author}{Oddo, P.}, \bibinfo{author}{Poulain, P.}, \bibinfo{author}{Falchetti, S.}, \bibinfo{author}{Storto, A.}, \bibinfo{author}{Zappa, G.}, \bibinfo{year}{2023}.
\newblock \bibinfo{title}{{Internal tides in the central Mediterranean Sea: observational evidence and numerical studies}}.
\newblock \bibinfo{journal}{Ocean Dyn.} \bibinfo{volume}{73}, \bibinfo{pages}{145--163}.
\newblock \DOIprefix\doi{10.1007/s10236-023-01545-z}.
\bibitem[{Pettenuzzo et~al.(2010)Pettenuzzo, Large and Pinardi}]{PLP10}
\bibinfo{author}{Pettenuzzo, D.}, \bibinfo{author}{Large, W.}, \bibinfo{author}{Pinardi, N.}, \bibinfo{year}{2010}.
\newblock \bibinfo{title}{{On the corrections of ERA-40 surface flux products consistent with the Mediterranean heat and water budgets and the connection between basin surface total heat flux and NAO}}.
\newblock \bibinfo{journal}{J. Geophys. Res.} \bibinfo{volume}{115}.
\newblock \DOIprefix\doi{10.1029/2009JC005631}.
\bibitem[{Raicich(1996)}]{R96}
\bibinfo{author}{Raicich, F.}, \bibinfo{year}{1996}.
\newblock \bibinfo{title}{{On the fresh balance of the Adriatic Sea}}.
\newblock \bibinfo{journal}{J. Mar. Sys.} \bibinfo{volume}{9}, \bibinfo{pages}{305--319}.
\newblock \DOIprefix\doi{10.1016/S0924-7963(96)00042-5}.
\bibitem[{Röske(2006)}]{R06}
\bibinfo{author}{Röske, F.}, \bibinfo{year}{2006}.
\newblock \bibinfo{title}{A global heat and freshwater forcing dataset for ocean models}.
\newblock \bibinfo{journal}{Ocean Modell.} \bibinfo{volume}{11}, \bibinfo{pages}{235--297}.
\newblock \DOIprefix\doi{10.1016/j.ocemod.2004.12.005}.
\bibitem[{Schwab and Rao(1983)}]{SR83}
\bibinfo{author}{Schwab, D.J.}, \bibinfo{author}{Rao, D.B.}, \bibinfo{year}{1983}.
\newblock \bibinfo{title}{{Barotropic oscillations of the Mediterranean and Adriatic Seas}}.
\newblock \bibinfo{journal}{Tellus A} \bibinfo{volume}{35A}, \bibinfo{pages}{417--427}.
\newblock \DOIprefix\doi{10.1111/j.1600-0870.1983.tb00216.x}.
\bibitem[{Shriver et~al.(2012)Shriver, Arbic, Richman, Ray, Metzger, Wallcraft and Timko}]{SAR12}
\bibinfo{author}{Shriver, J.F.}, \bibinfo{author}{Arbic, B.K.}, \bibinfo{author}{Richman, J.G.}, \bibinfo{author}{Ray, R.D.}, \bibinfo{author}{Metzger, E.J.}, \bibinfo{author}{Wallcraft, A.J.}, \bibinfo{author}{Timko, P.G.}, \bibinfo{year}{2012}.
\newblock \bibinfo{title}{{An evaluation of the barotropic and internal tides in a high-resolution global ocean circulation model}}.
\newblock \bibinfo{journal}{J. Geophys. Res.} \bibinfo{volume}{117}.
\newblock \DOIprefix\doi{10.1029/2012JC008170}.
\bibitem[{St.~Laurent et~al.(2002)St.~Laurent, Simmons and Jayne}]{SSJ02}
\bibinfo{author}{St.~Laurent, L.C.}, \bibinfo{author}{Simmons, H.L.}, \bibinfo{author}{Jayne, S.R.}, \bibinfo{year}{2002}.
\newblock \bibinfo{title}{{Estimating tidally driven mixing in the deep ocean}}.
\newblock \bibinfo{journal}{Geophys. Res. Lett} \bibinfo{volume}{29}.
\newblock \DOIprefix\doi{10.1029/2002GL015633}.
\bibitem[{von Storch et~al.(2023)von Storch, Hertwig, Lüschow, Brüggemann, Haak, Korn and Singh}]{vSHL23}
\bibinfo{author}{von Storch, J.S.}, \bibinfo{author}{Hertwig, E.}, \bibinfo{author}{Lüschow, V.}, \bibinfo{author}{Brüggemann, N.}, \bibinfo{author}{Haak, H.}, \bibinfo{author}{Korn, P.}, \bibinfo{author}{Singh, V.}, \bibinfo{year}{2023}.
\newblock \bibinfo{title}{{Open-ocean tides simulated by ICON-O, version icon-2.6.6}}.
\newblock \bibinfo{journal}{Geosci. Model Dev.} \bibinfo{volume}{16}, \bibinfo{pages}{5179--5196}.
\newblock \DOIprefix\doi{10.5194/gmd-16-5179-2023}.
\bibitem[{Tsimplis et~al.(1995)Tsimplis, Proctor and Flathe}]{TPF95}
\bibinfo{author}{Tsimplis, M.N.}, \bibinfo{author}{Proctor, R.}, \bibinfo{author}{Flathe, R.A.}, \bibinfo{year}{1995}.
\newblock \bibinfo{title}{{A two-dimensional tidal model for the Mediterranean Sea}}.
\newblock \bibinfo{journal}{J. Geophys. Res.} \bibinfo{volume}{100}, \bibinfo{pages}{16223--16239}.
\newblock \DOIprefix\doi{10.1029/95JC01671}.

\end{thebibliography}
\pagebreak

\appendix

\section{Spectral analysis}
\label{spectra}

In this work, spectra are calculated to analyse the frequencies and wavelengths at which internal tides are generated and propagated at a variety of temporal and spatial scales. In the following sections, each type of spectrum is briefly defined. 

\subsection{Spectrum calculation}

For all spectra, a periodogram is used as an estimate of the spectrum. The main advantages of using the periodogram over other estimators of a spectrum is that a periodogram is asymptotically unbiased, and estimates at adjacent frequencies are almost uncorrelated. The periodogram is calculated as follows: 

For a timeseries ${x_1, ..., x_T}$, where $x_t, t = 1, ..., T$ is periodic and can be expanded in terms of sine and cosine functions, $x_t$ is;

\begin{equation}
    x_t = a_0 + \sum_{j=1}^q (a_j \cos(2\pi \omega_j t) + b_j \sin(2\pi \omega_j t)),
    \omega_j = \frac{j}{T}, 
    j = 1, ..., q, 
    q = \frac{T^*}{2}  
\end{equation}

where $T^*$ is the largest integer within $T/2$, and 

\begin{equation}
    a_0 = \frac{1}{T} \sum_{t=1}^T x_t
\end{equation}

\begin{equation}
    a_j = \frac{2}{T} \sum_{t=1}^T x_t \cos(2\pi \omega_j t)
\end{equation}

\begin{equation}
    b_j = \frac{2}{T} \sum_{t=1}^T x_t \sin(2\pi \omega_j t)
\end{equation}

The periodogram at the frequency $\omega_j$ is then:

\begin{equation}
    I_{T_j} = \frac{T}{4}(a_j^2 + b_j^2), j = \frac{T^*-1}{2}, ..., \frac{T^*}{2}.
\end{equation}

This periodogram is used to estimate power spectra from timeseries at single points in the ocean. 

\subsection{Wavenumber spectrum}

In addition to spectra in frequency space, wavenumber spectra are calculated to find the wavelengths of the internal tides. These spectra are calculated in wavenumber space, along a length of physical space at a specific time, rather than in frequency space using a time series of data. The periodogram is also used for this calculation, replacing the frequency $\omega_j$ with the wavenumber $k_j$ and the time period $T$ with wavelength $\lambda$. A harmonic analysis of currents for each tidal component is used to calculate the periodogram, to analyse the wavelengths at tidal frequencies.

\section{Model intercomparison}
\label{intercomparison}
\setcounter{figure}{0}
\setcounter{table}{0}
\setcounter{subsection}{0}

The generation, distribution, and propagation routes of internal tides in the Mediterranean Sea in NEMO and ICON experiment outputs are compared to one another in Section \ref{results}. Although the two experiments give broadly similar results, they differ in the exact generation sites and particularly in the wavelengths of the first modes of the M2 and K1 internal tides. One reason for this could be different representation of barotropic lunisolar tides in the two models. Here we briefly compare the barotropic tides in the two experiments.

\begin{figure}[h]
    \centering
    \includegraphics[width=\linewidth]{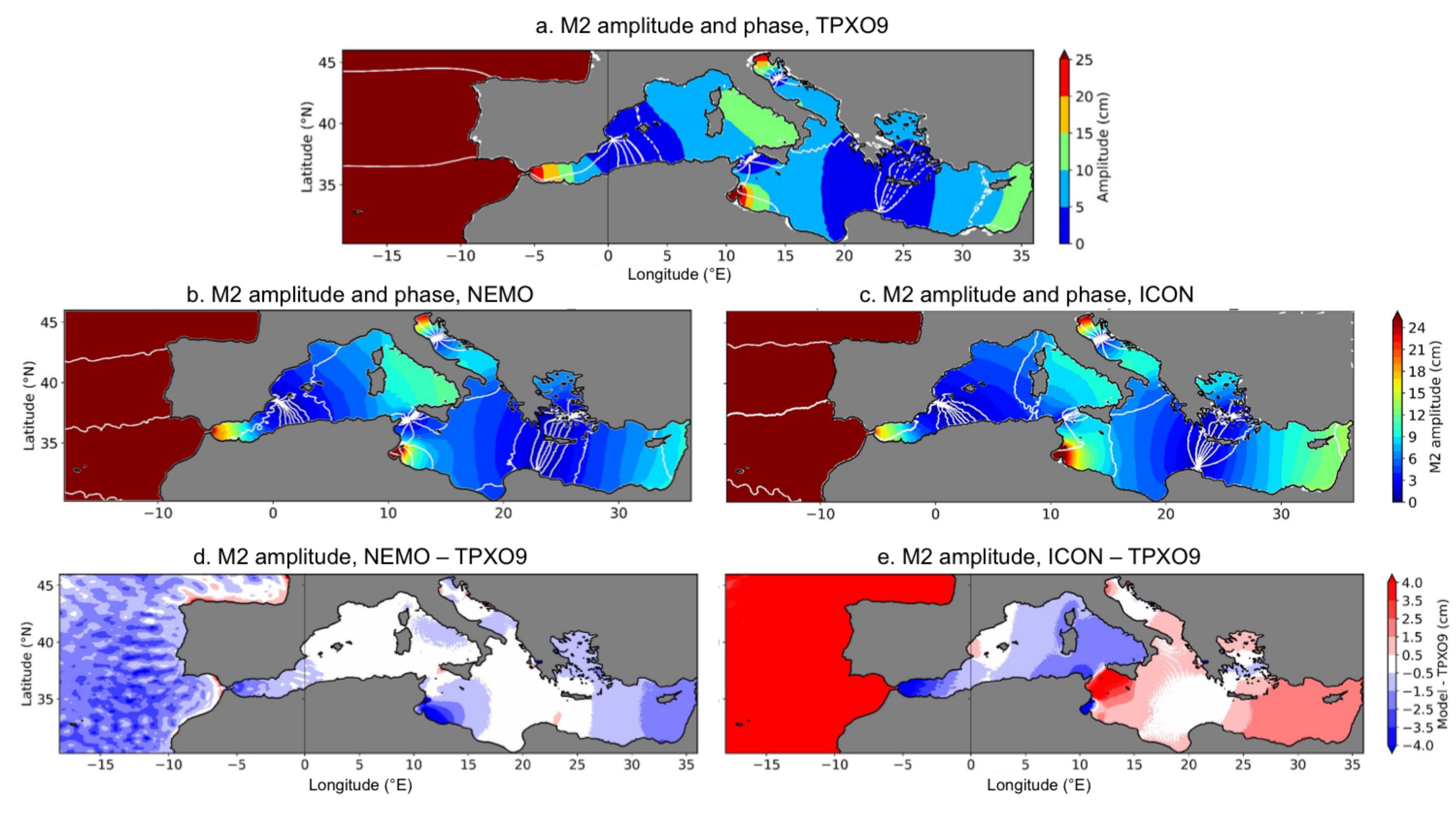}
    \caption{Amplitude and phase of M2 tidal component in a. TPXO9, b. NEMO, and c. ICON for March 2022, and the amplitude difference between TPXO9 and d. NEMO and e. ICON.}
    \label{fig:validation_m2}
\end{figure}

\begin{figure}[h]
    \centering
    \includegraphics[width=\linewidth]{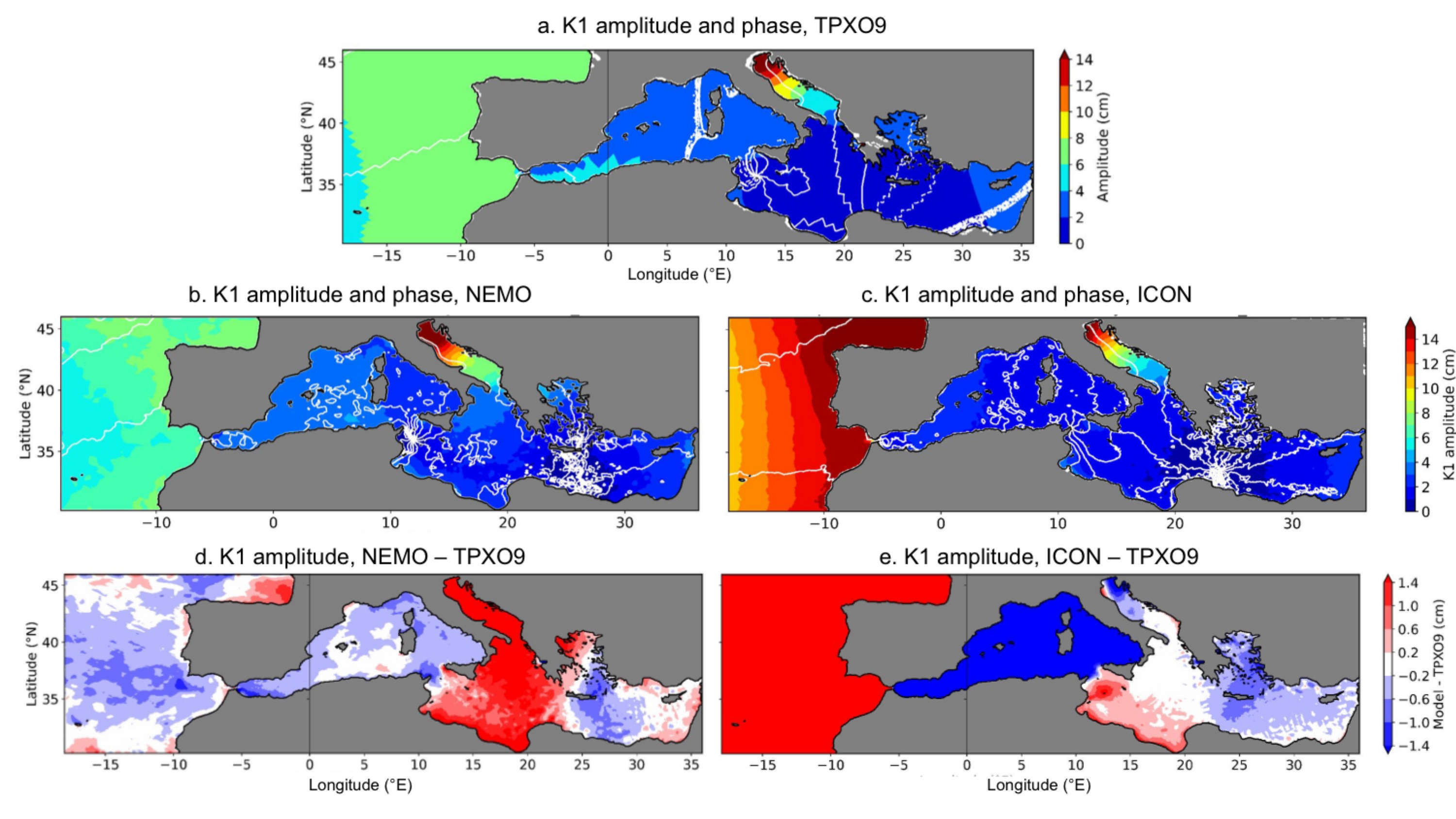}
    \caption{Amplitude and phase of K1 tidal component in a. TPXO9, b. NEMO, and c. ICON for March 2022, and the amplitude difference between TPXO9 and d. NEMO and e. ICON.}
    \label{fig:validation_k1}
\end{figure}

The tidal amplitudes and phases were calculated through a harmonic analysis of sea surface height and compared to data from TPXO9 \cite[]{ER03}. Amplitudes and phases, and the vector difference between each experiment and TPXO9, for the principal semidiurnal and diurnal tidal components (M2 and K1) are shown in Figures \ref{fig:validation_m2} and \ref{fig:validation_k1}. For the analysed month, placements of the maximum amplitudes and amphidromic points are broadly correct in both the experiments within the Mediterranean Sea. However, it is clear that the models do not always correctly capture the tidal amplitude in key locations for internal tide generation. In the ICON implementation, areas close to the Gibraltar and Sicily Straits have particularly large differences from TPXO9 for the M2 component, and for the K1 component, the tidal amplitude is greatly underestimated in the western Mediterranean basin. In the NEMO experiment, the underestimation of K1 amplitude is limited to fewer areas, but notably is overestimated in the north Adriatic Sea, where K1 is typically large. The NEMO experiment represents the M2 component fairly well, with underestimation of tidal amplitude limited to a few regions. 

\begin{figure}
    \centering
    \includegraphics[width=\linewidth, trim=0cm 0cm 0cm 1cm, clip]{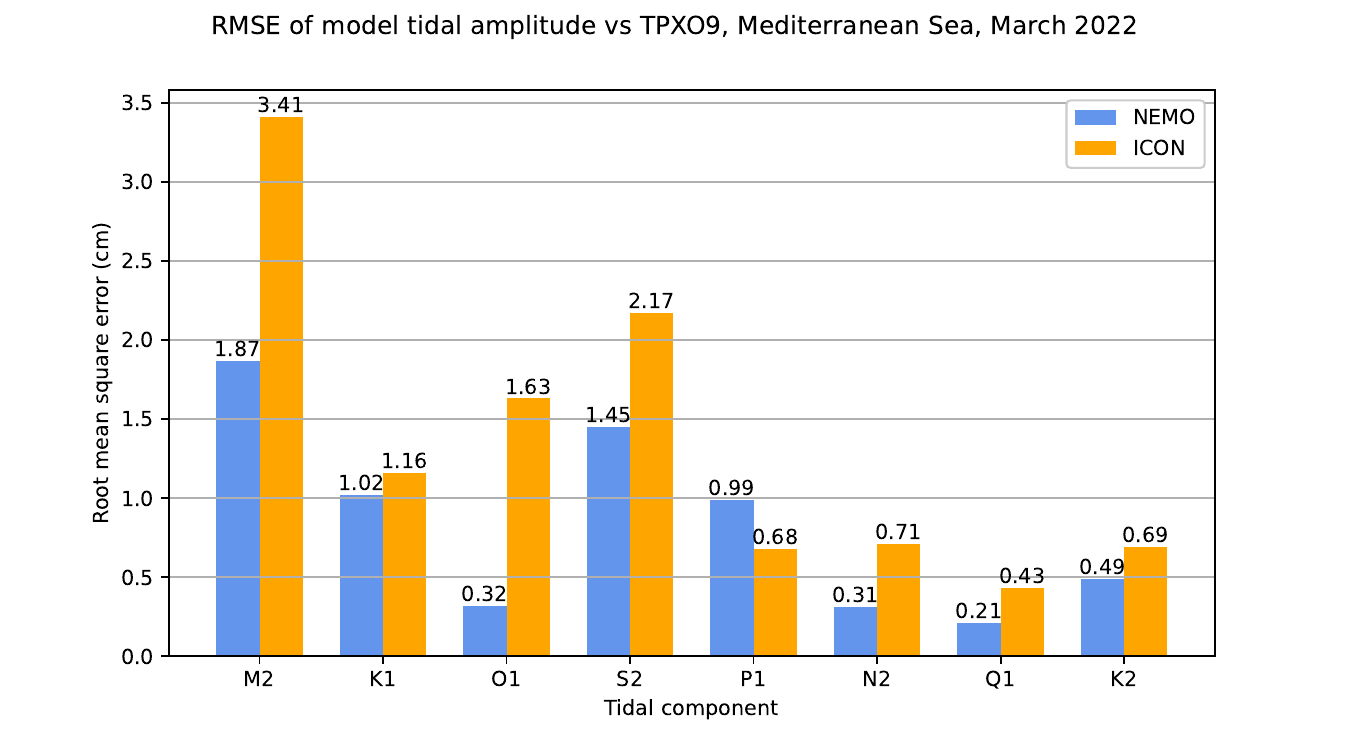}
    \caption{Root mean square error of model tidal amplitude for eight tidal components compared to TPXO9 in the Mediterranean Sea for March 2022. NEMO is in blue and ICON is shown in orange.}
    \label{fig:rmse_tides_NEMO-ICON}
\end{figure}

Since the NEMO model includes eight tidal components, the root mean square error of the NEMO and ICON tidal amplitudes compared to TPXO9 were calculated for these eight components and are summarised in Figure \ref{fig:rmse_tides_NEMO-ICON}. NEMO has lower root mean square errors than ICON for all components other than P1.

\clearpage
\FloatBarrier

\section{Additional regional analysis}
\label{extraregions}
\setcounter{figure}{0}
\setcounter{table}{0}
\setcounter{subsection}{0}

In Section \ref{results}, spectra of horizontal currents in three points: the Gibraltar Strait, Sicily Strait, and Ionian Sea (north), are shown. Spectra of the remaining points from Figure \ref{fig:domain_map} are in the following figures.

\subsection{Aegean Sea}
\begin{figure}[h]
\centering
\includegraphics[width=\linewidth, trim=0cm 0cm 0cm 0cm, clip]{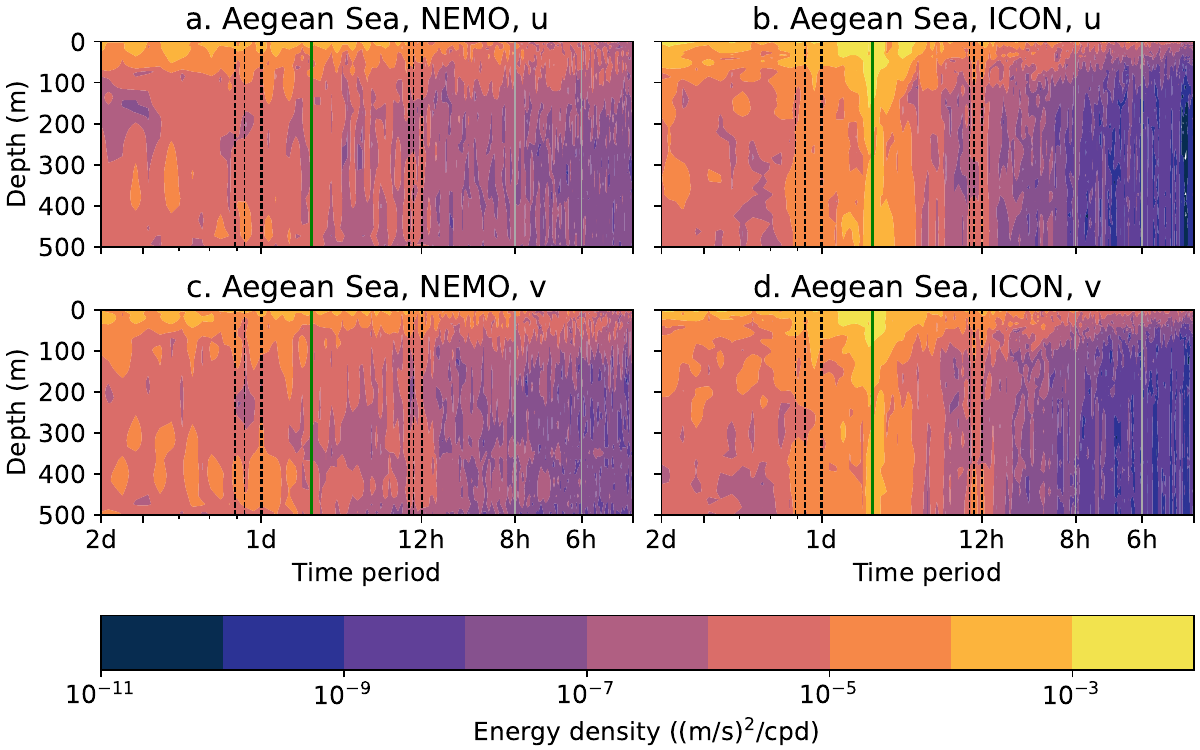}
\caption{Spectra of baroclinic zonal and meridional currents in the Aegean Sea (38.44$^\circ$N, 25.29$^\circ$E), for a. zonal currents in NEMO, b. zonal currents in ICON, c. meridional currents in NEMO, and d. meridional currents in ICON, through all depths. Green vertical line is the inertial frequency at this latitude.}
\label{fig:rotary_aegean}
\end{figure}

\clearpage
\FloatBarrier

\subsection{Algerian Sea}
\begin{figure}[h]
\centering
\includegraphics[width=\linewidth, trim=0cm 0cm 0cm 0cm, clip]{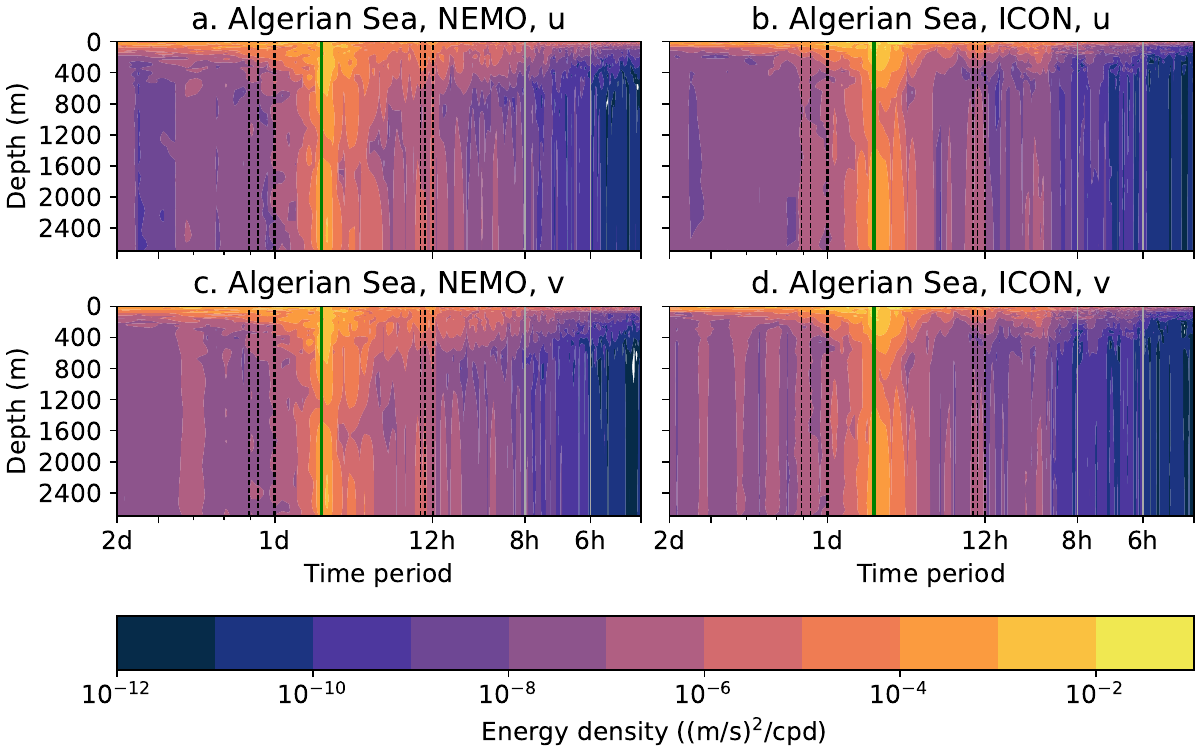}
\caption{Spectra of baroclinic zonal and meridional currents in the Algerian Sea (37.85$^\circ$N, 4.73$^\circ$E), for a. zonal currents in NEMO, b. zonal currents in ICON, c. meridional currents in NEMO, and d. meridional currents in ICON, through all depths. Green vertical line is the inertial frequency at this latitude.}
\label{fig:rotary_algerian}
\end{figure}

\clearpage
\FloatBarrier

\subsection{Ionian Sea (South)}
\begin{figure}[h]
\centering
\includegraphics[width=\linewidth, trim=0cm 0cm 0cm 0cm, clip]{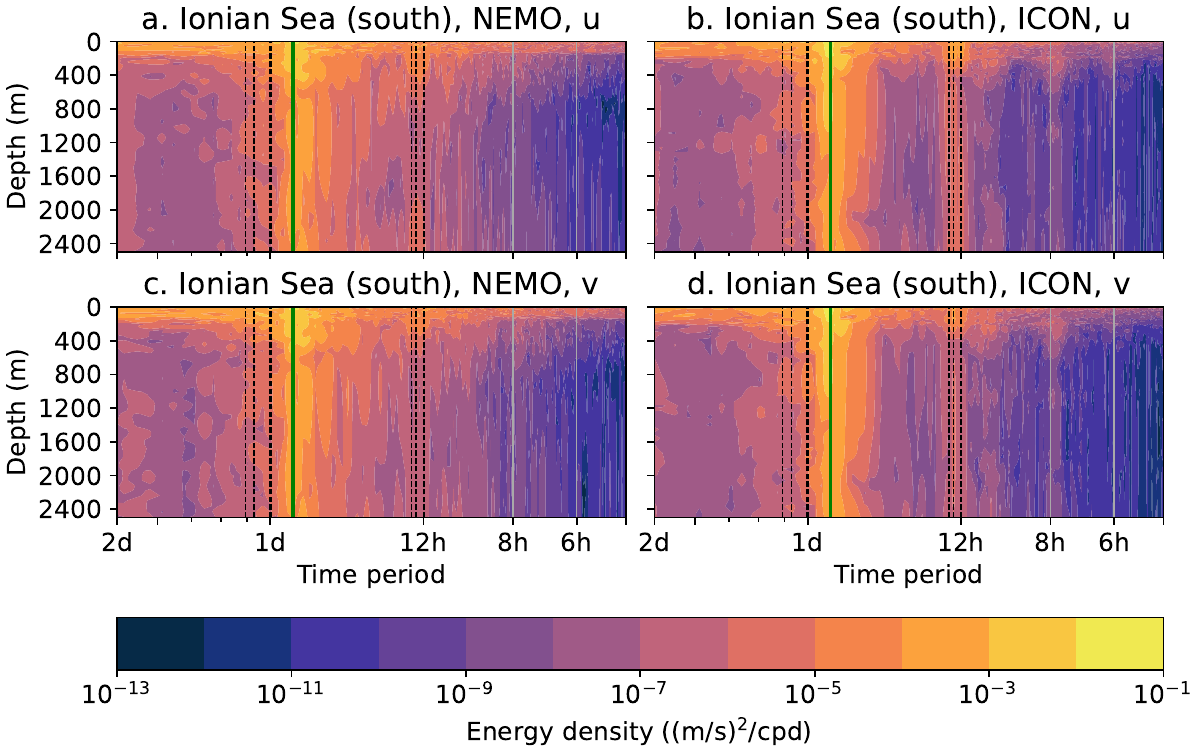}
\caption{Spectra of baroclinic zonal and meridional currents in the Ionian Sea (33.65$^\circ$N, 19.48$^\circ$E), for a. zonal currents in NEMO, b. zonal currents in ICON, c. meridional currents in NEMO, and d. meridional currents in ICON, through all depths. Green vertical line is the inertial frequency at this latitude.}
\label{fig:rotary_libya}
\end{figure}

\clearpage
\FloatBarrier

\subsection{Malta Bank}
\begin{figure}[h]
\centering
\includegraphics[width=\linewidth, trim=0cm 0cm 0cm 0cm, clip]{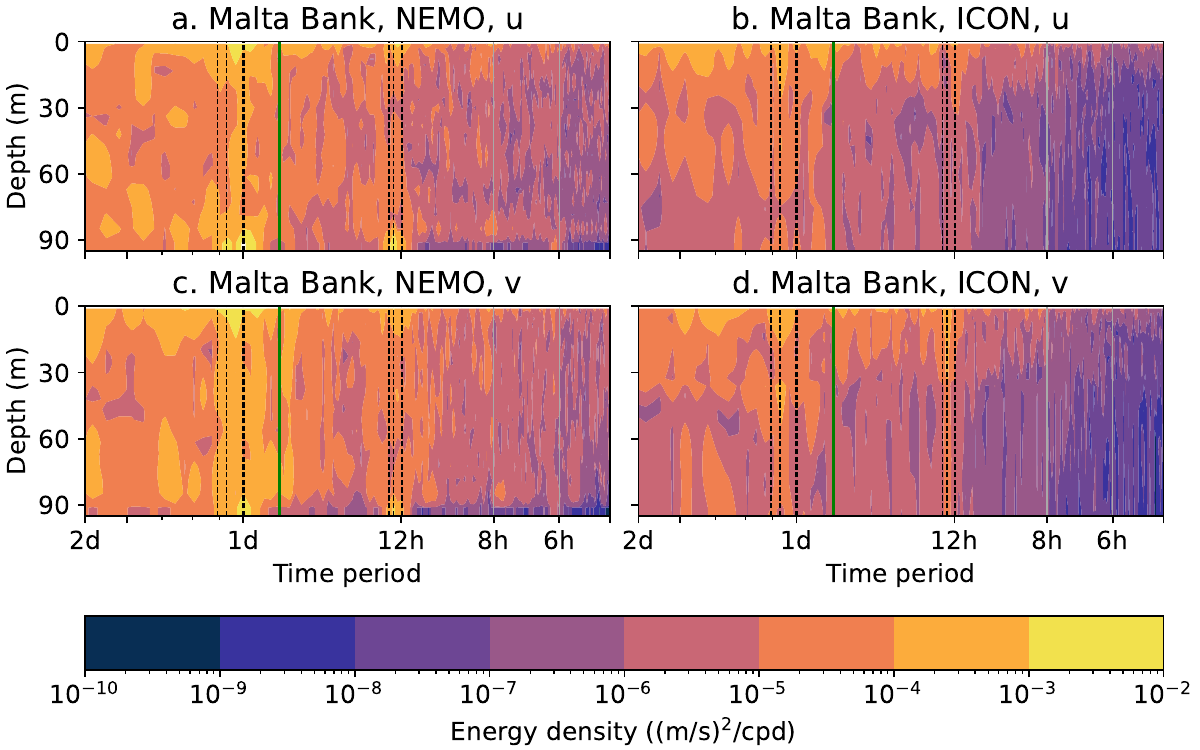}
\caption{Spectra of baroclinic zonal and meridional currents in the Malta Bank (35.98$^\circ$N, 14.50$^\circ$E), for a. zonal currents in NEMO, b. zonal currents in ICON, c. meridional currents in NEMO, and d. meridional currents in ICON, through all depths. Green vertical line is the inertial frequency at this latitude.}
\label{fig:rotary_malta_bank}
\end{figure}

\clearpage
\FloatBarrier

\subsection{Tyrrhenian Sea}
\begin{figure}[h]
\centering
\includegraphics[width=\linewidth, trim=0cm 0cm 0cm 0cm, clip]{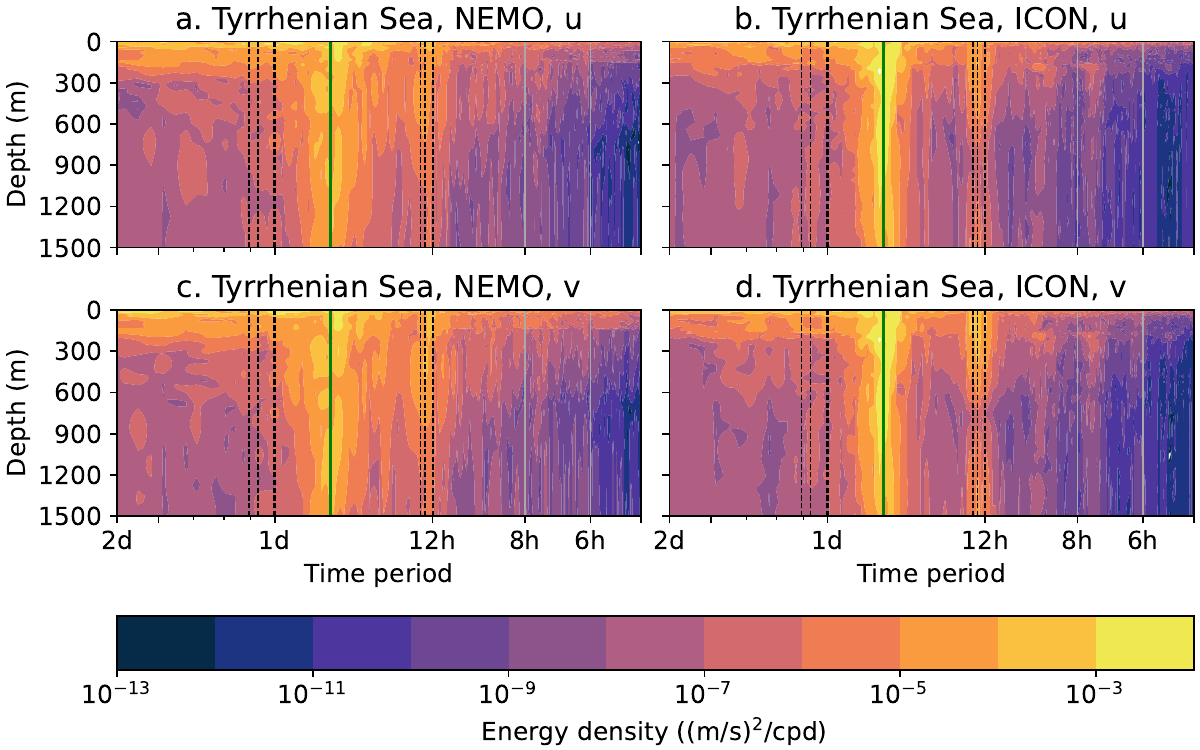}
\caption{Spectra of baroclinic zonal and meridional currents in the Tyrrhenian Sea (39.73$^\circ$N, 11.44$^\circ$E), for a. zonal currents in NEMO, b. zonal currents in ICON, c. meridional currents in NEMO, and d. meridional currents in ICON, through all depths. Green vertical line is the inertial frequency at this latitude.}
\label{fig:rotary_tyrrhenian}
\end{figure}

\section{The Sturm-Liouville eigenvalue problem}
\label{sturmliouville}
\setcounter{figure}{0}
\setcounter{table}{0}
\setcounter{subsection}{0}
\setcounter{equation}{0}
Solutions to the Sturm-Liouville eigenvalue problem allow us to calculate the wavelengths of internal tides to compare to the low mode internal tide wavelengths found through wavenumber analysis (see Section 3.3) and confirm that the peaks found through wavenumber analysis are indeed due to internal tides. The Sturm-Liouville problem uses the squared Brunt-Väisälä frequency from the model outputs, interpolated onto a regular vertical grid with a 1m resolution. The eigenvalue problem is defined as:

\begin{equation}
    \frac{1}{N^2} \frac{d^2}{dz^2}w_m(z) = -v_m w_m, m = 1, 2, ...
\end{equation}

where $z$ represents the vertical axis, $N^2$ is the squared Brunt-Väisälä frequency, $w_m$ is the eigenvector, and $v_m$ is the eigenvalue of the $m$th vertical mode. The eigenvalues are then used to calculate the wavelength of the internal tide, $L_m$ by:

\begin{equation}
    L_m = \frac{1}{\sqrt{v_m(\omega^2 - f^2)}}
\end{equation}

where $\omega$ is the frequency of the internal tide, and $f$ is the Coriolis parameter. 

There are several assumptions that are made when using this method to calculate the wavelength of internal tides, primarily that a flat bottom is assumed. Due to the high resolution of the models used in this work, this assumption is significant and is a caveat to this analysis. Moreover, we also neglect any possible variations in stratification and internal tide wavelength within a selected region, which is likely to have varying importance depending on the region and the depth. 

\end{document}